\documentclass[10pt, a4paper]{book}
\usepackage{amsmath}
\usepackage{amsfonts}
\usepackage{fullpage}
\usepackage{hyperref}
\usepackage{comment}
\usepackage[dvipsnames, table]{xcolor}
\usepackage{graphicx}
\usepackage{array}
\usepackage{makecell}
\usepackage{tabularx}
\usepackage{multirow}
\usepackage{nicematrix}
\usepackage{soul}
\usepackage{enumitem,calc}
\usepackage[skins]{tcolorbox}
\usepackage{float}
\usepackage{enumitem}
\usepackage{forest}
\usepackage{algorithm}
\usepackage{algpseudocode}
\usepackage{subfiles}
\usepackage{amssymb}
\usepackage{fontawesome5}
\usepackage{pgf-umlsd}
\usepackage{array}
\usepackage{float}
\usetikzlibrary{arrows.meta, positioning, calc, shapes.geometric, fit}
\usepackage[
backend=biber,
sorting=ynt
]{biblatex}
\graphicspath{{images/}}

\sethlcolor{gray!20}

\def\bitcoin{%
  \leavevmode
  \vtop{\offinterlineskip %\bfseries
    \setbox0=\hbox{B}%
    \setbox2=\hbox to\wd0{\hfil\hskip-.03em
    \vrule height .3ex width .15ex\hskip .08em
    \vrule height .3ex width .15ex\hfil}
    \vbox{\copy2\box0}\box2}}

\hypersetup{
colorlinks=false,
pdftitle={Wrapless-20Book},
pdfpagemode=FullScreen,
}
\newtcolorbox{scriptPubKey}{
  enhanced,
  boxsep=1pt,
  arc=0.75ex,
  colback=gray!10,
  colframe=gray!40,
  boxrule=1pt,
  leftrule=80pt,
  top=-3.5mm,
  overlay unbroken and first ={%
    \node[minimum width=1cm,
      anchor=south,
      font=\sffamily\bfseries,
      xshift=40pt,
      yshift=-6.5pt,
    black]
    at (frame.west) {scriptPubKey:};
  }
}
\newtcolorbox{redeemScript}{
  enhanced,
  boxsep=1pt,
  arc=0.75ex,
  colback=gray!10,
  colframe=gray!40,
  boxrule=1pt,
  leftrule=80pt,
  top=-3.5mm,
  overlay unbroken and first ={%
    \node[minimum width=1cm,
      anchor=south,
      font=\sffamily\bfseries,
      xshift=40pt,
      yshift=-6.5pt,
    black]
    at (frame.west) {redeemScript:};
  }
}
\newtcolorbox{scriptSig}{
  enhanced,
  boxsep=1pt,
  arc=0.75ex,
  colback=gray!10,
  colframe=gray!40,
  boxrule=1pt,
  leftrule=80pt,
  top=-3.5mm,
  overlay unbroken and first ={%
    \node[minimum width=1cm,
      anchor=south,
      font=\sffamily\bfseries,
      xshift=40pt,
      yshift=-6.5pt,
    black]
    at (frame.west) {scriptSig:};
  }
}

\newtcolorbox{ideablock}[2][]{
    colback=white,
    colframe=black,
    fonttitle=\bfseries,
    title={\faIcon{lightbulb} \quad #2},
    #1
}

\newcommand{\bscomment}[1]{// & \text{ #1}}
\newcommand{\elem}[1]{\, \langle #1 \rangle \,}
\newcommand{\opcode}[1]{\, \texttt{#1} \,}

\newcolumntype{C}[1]{>{\centering\arraybackslash}m{#1}}

\title{Wrapless book \\ \small v.0.4}
\author{Rarimo Protocol}
\date{\today}

\begin{document}
\maketitle
\tableofcontents
\newpage

\section*{Notation}
Let $\mathbb{G}$ be a cyclic group of prime order $p$ written additively, $G \in \mathbb{G}$ is the group generator. We define $a \in \mathbb{F}_p$ as a scalar value and $A \in \mathbb{G}$ as a group element. In this paper, we are referring to the ECC group with $\mathsf{secp256k1}$ curve and all parameters defined in its specification \cite{Telephone2008SECXR}.

We define a hash function as $\mathsf{H_{alg}}: \{0,1\}^* \rightarrow \{0,1\}^{n}$, where $\mathsf{alg}$ is the used algorithm (usually we will refer to SHA-256 with $n=256$ and RIPEMD-160 with $n=160$). Additionally, we use a separate hash function with the field element output $\mathsf{H_{alg_F}}: \{0,1\}^* \rightarrow \mathbb{F}_p$ (Poseidon \cite{poseidon}, for instance).

Let $\mathsf{sigGen_{alg}}:(m,sk)\to\mathsf{\sigma}$ be signature generation function and $\mathsf{sigVer_{alg}}:(m,P, \sigma)\to\{0,1\}$ a signature verification function, where $\mathsf{alg}$ is a algorithm used, $m \in \mathbb{B}^*$ is the message, $sk \in \mathbb{F}_p$ and $P=sk\cdot G \in \mathbb{G}$ is a signer's key pair.

We define the relation for the proof $\pi$ as $\mathcal{R} = \{(w;x) \in \mathcal{W} \times \mathcal{X}: \phi_1(w,x), \phi_2(w,x) , \dots, \phi_m(w,x)\}$, where $w$ is a witness data, $x$ is a public data and $\phi_1(w,x), \phi_2(w,x) , \dots, \phi_m(w,x)$ the set of relations must be proven simultaneously.

We will use a short notation for Bitcoin transactions as:

\[
\mathsf{TX}\{(\mathsf{id, i, proof})^{(n)};(\mathsf{a \bitcoin, cond})^{(m)}\}
\]

with $n$ inputs and $m$ outputs, where $\mathsf{id}$ is the reference to the previous transaction, $\mathsf{i}$ -- a corresponding output's index, $\mathsf{proof}$ -- the list of data which is needed for transaction spending, $\mathsf{a}$ -- the number of coins in the output, $\mathsf{cond}$ -- scriptPubKey conditions. 

To visualize the transaction details, we will use the following notation:\\

\begin{tabular}{m{4cm}|m{4cm}|m{6cm}}
  \textbf{Inputs} & \textbf{Outputs} & \textbf{Output Script} \\
  \hline
  \texttt{OP\_0 <$\sigma_A$> <$\sigma_B$>} \small{(2-of-2 multisignature)} & $a_0\bitcoin$  & \texttt{OP\_DUP OP\_HASH160 OP\_PUSHBYTES\_20 <addr\_A> OP\_EQUALVERIFY OP\_CHECKSIG} \small{(P2PKH with Alice's address)}\\
  \hline
  & $b_0\bitcoin$ & \texttt{OP\_HASH160 OP\_PUSHBYTES\_20 <hash> OP\_EQUAL} \small{(P2SH with conditions commited by the hash)} \\
\end{tabular} \\

For detailed visualization of spending conditions in Bitcoin Script \cite{bitcoin-script}, we would show colored boxes like these:

\begin{scriptSig}
\begin{align*}
& \elem{\mathsf{\sigma_a}} & \bscomment{Push a signatures on stack}
\end{align*}
\end{scriptSig}

\begin{scriptPubKey}
\begin{align*}
& \elem{\mathsf{P_a}} & \bscomment{Push a public key on stack} \\
& \opcode{OP\_CHECKSIG} & \bscomment{Bitcoin Script opcode}
\end{align*}
\end{scriptPubKey}

Where \textbf{scriptPubKey} is the previous output's spending condition (also mentioned above as $\mathsf{cond}$) and \textbf{scriptSig} (also $\mathsf{proof}$) are values (or script) that satisfy a spending condition from \textbf{scriptPubKey}, in the current transaction input.

\begin{ideablock}{Idea Block}
    In blocks like this, we will include some side notes that are required for a deeper understanding of the further conclusions.
\end{ideablock}
\newpage

\chapter{Introduction}
There is only one action people can take with bitcoins in a trustless manner: transfer. Bitcoin's programmability is limited, which does not allow us to launch popular DeFi protocols that operate with native BTC. Therefore, the popular approach is to wrap bitcoins and utilize them within the existing DeFi infrastructure. The problem is that the existing wrapping mechanisms are either centralized or federated with m-of-n trust assumptions.

One way to program BTC behavior is through DLC contracts \cite{dlc-contracts}. This method relies on the oracle, which, signing different data, can define the coins' spending path. The primary drawback of this approach is that the oracle serves as the trusted entity. Despite the oracle's inability to steal funds, it can manipulate the decision acceptance in favor of one of the parties.

There is massive potential in BitVM2 technology \cite{bitvm2}. Still, it requires some updates to Bitcoin, such as the addition of \texttt{OP\_CTV} \cite{bip119} + \texttt{OP\_CSFS} \cite{bip348} or \texttt{OP\_CAT} \cite{bip420}, to achieve a more reliable security model (better than 1-of-n) and decrease the operating cost.

This paper aims to design \textbf{Wrapless} -- a lending protocol that enables the collateralization of BTC without requiring a trusted wrapping mechanism. This paper describes a solution that enables locking BTC in the "loan channel" and providing a loan on another blockchain platform, such that the loan's installment (which can be partial) results in unlocking the corresponding amount of BTC.

The advantage of Wrapless is that it doesn't rely on any trusted party and represents a game between a borrower and a lender, where cheating by any party is economically irrational. The most significant limitation we observe is that the solution is peer-to-peer and does not facilitate liquidity aggregation.

\chapter{Preliminaries}
Before diving into the Wrapless protocol, we recommend reviewing the technologies it is based on. If you are familiar with them, you can skip this section.

\section{Merkle Trees}

\subsection{Merkle Tree Construction}
A \emph{Merkle Tree} \cite{merkle1988} is a complete binary tree built from a sequence of data blocks, where each leaf encodes the hash of a block and each internal node is computed as the hash of the concatenation of its children. This construction enables efficient inclusion proofs with logarithmic complexity.

In Bitcoin, Merkle trees are used to commit to the list of transactions in a block. The Merkle root is included in the block header, allowing Simplified Payment Verification (SPV) clients to verify transaction inclusion using short proofs, without the need to download the full block.
The Merkle tree is constructed as follows:
\begin{center}
\begin{minipage}{0.9\linewidth}
\begin{algorithm}[H]
\caption{Merkle Tree Construction}
\begin{algorithmic}[1]
\Require Transactions $x_1, \dots, x_n$
\Ensure Merkle root $y$
\State $L \gets [\mathsf{H}(x_1), \mathsf{H}(x_2), \dots, \mathsf{H}(x_n)]$
\While{$|L| > 1$}
    \If{$|L|$ is odd}
        \State Append $L[-1]$ to $L$ \Comment{Duplicate last element}
    \EndIf
    \State $L \gets [ \mathsf{H}(L_{2i} \Vert L_{2i+1}) \;|\; i = 0,\dots,|L|/2 - 1 ]$
\EndWhile
\State \Return $L[0]$
\end{algorithmic}
\end{algorithm}
\end{minipage}
\end{center}

The Merkle root $y$ serves as a commitment to the entire set of transactions. A \emph{Merkle proof} consists of $\log_2 n$ sibling hashes and is sufficient to recompute the path to the root. Given a transaction $x_i$, Merkle root $y'$ and a proof $\pi_i$ for the leaf index $i$, the verifier:
\begin{enumerate}
    \item Computes $\mathsf{H}(x_i)$.
    \item Iteratively hashes with the siblings in $\pi_i$ up to the root $y$.
    \item Accepts if the result equals $y'$.
\end{enumerate}
A classic Merkle tree construction supports \emph{inclusion proofs}, but does not provide efficient non-inclusion proofs since the structure does not encode which leaves are assigned. For non-inclusion proofs, we can use the SMT construction described below.

\subsection{Sparse Merkle Tree}
A \emph{Sparse Merkle Tree} (SMT) \cite{smt} is a fixed-depth binary tree used to commit to a large but sparsely populated key-value map efficiently. Unlike classical Merkle trees, SMTs allow for efficient proofs of both inclusion and non-inclusion. They are particularly useful for storing finalized blockchain state, such as account balances or rollup commitments, where only a small subset of the entire key space is populated.

We define a default value $\bot$ representing empty leaves. The SMT is constructed by inserting known $(k, v)$ pairs into an otherwise empty tree of depth $k$.

\begin{center}
\begin{minipage}{0.9\linewidth}
\begin{algorithm}[H]
\caption{Sparse Merkle Tree Construction}
\begin{algorithmic}[1]
\Require Map $M : \mathcal{K} \to \mathcal{V}$, depth $k$
\Ensure Root hash $r$
\State Initialize all $2^k$ leaves to $h(\bot)$
\For{each $(key, value) \in M$}
    \State $p \gets \text{binary path of } key$
    \State $h_p \gets \mathsf{H}(value)$
    \For{$i = k-1$ down to $0$}
        \State Combine child hashes to compute parent
    \EndFor
\EndFor
\State \Return root $r$
\end{algorithmic}
\end{algorithm}
\end{minipage}
\end{center}

Each proof consists of the $k$ sibling hashes along the path from a leaf to the root. For non-inclusion, the proof shows that the leaf is unassigned and the hash chain is consistent with default values. Given a root $r$ and a proof $\pi$ for key $k$, the verifier:
\begin{enumerate}
    \item Computes the binary path of $k$ and the expected hash of its value (or default).
    \item Hashes upward using the siblings in $\pi$ to reconstruct the root.
    \item Accepts if the result equals $r$.
\end{enumerate}
Because every leaf is implicitly defined (either with real data or a default), SMTs enable efficient proofs of both \emph{inclusion} and \emph{non-inclusion}.

\section{SPV Contract}
The SPV (Simplified Payment Verification) node concept, initially proposed in the Bitcoin whitepaper \cite{bitcoin}, allows the verification of Bitcoin transactions without the need to maintain a full node. Instead of downloading the entire blockchain, the SPV node only synchronizes block headers and their corresponding Merkle roots. This SPV approach enables the verification of transaction inclusion by performing Merkle proof verification, which relies on the integrity of the Merkle root stored in the block header.

It is possible to launch the SPV node as a smart contract. This contract allows a trustless synchronization of the entire Bitcoin history to the required blockchain. Such a property is quite helpful for cross-chain applications.

\subsection{Block Header Validation Rules}
When the SPV node is syncing with the Bitcoin network, it receives the block headers and verifies them according to the set of rules:
\begin{enumerate}
    \item Structure validity and existence of all fields:
    \begin{enumerate}[label=\roman*)]
        \item Version (4 bytes), Previous block (32 bytes), Merkle Root (32 bytes), Timestamp (4 bytes), Bits (4 bytes), Nonce (4 bytes).
    \end{enumerate}
    
    \item Previous block hash correctness: 
    \begin{enumerate}[label=\roman*)]
        \item Verifies that the block, the header of which was received, is part of the mainchain and refers to the existing previous block.
    \end{enumerate}
    
    \item Timestamp: 
    \begin{enumerate}[label=\roman*)]
        \item Verifies if the timestamp value exceeds the median value of the previous 11 blocks. Additionally, the node does not accept blocks with timestamps more than two hours in the future.
    \end{enumerate}
    
    \item PoW verification: 
    \begin{enumerate}[label=\roman*)]
        \item The block’s header double hash (SHA256) value must satisfy the defined difficulty parameter (must be less than the target value).
        \item The difficulty target parameter is changed every 2016 blocks to adjust the block mining time for the current network hash rate. It does so by summing up the total number of minutes miners took to mine the last 2,015 blocks and dividing this number by the protocol’s desired goal of 20,160 minutes (2,016 blocks x 10 minutes). The ratio is then multiplied by the current difficulty level to produce the new difficulty level.
        \item If the correction factor is greater than 4 (or less than 1/4), then 4 or 1/4 is used instead to prevent abrupt changes.
    \end{enumerate}
    
    \item Nonce validation: 
    \begin{enumerate}[label=\roman*)]
    \item The hash value of the concatenation of all previous parts of the header with the nonce value must be equal to the block hash value (satisfy difficulty target parameter).
    \end{enumerate}
\end{enumerate}
    
\subsection{Transaction Inclusion Verification}
For transaction verification, the SPV node requests the full node to return the proof of transaction inclusion in the block. As proof, in this case, the Merkle branch is being used, which must lead to the existing Merkle root, defined in one of the mainchain block headers. We will use the notation $\pi_{\mathsf{SPV}}(\mathsf{TX})$ as an SPV proof that includes a block hash and an inclusion proof of transaction $\mathsf{TX}$.

\subsection{Reorganizations Management}
When verifying the existence of a Bitcoin transaction, the SPV node must ensure that the block is included in the mainchain, which is the heaviest chain.

Finality in Bitcoin is probabilistic, which means the chain can always be reorganized. The SPV node must be able to switch to the alternative main chain if its total difficulty is greater. 

\subsection{Proposed SPV Contracts Architecture}
The SPV contract serves as a storage for the Bitcoin block history. Similar to the SPV node, the SPV contract stores block headers provided by users and verifies their validity through a series of checks. Once a block header is successfully verified, it is stored within the contract, allowing other users to access the data with confidence in its correctness. The contract implements read methods to facilitate data retrieval, thereby ensuring a reliable source of verified Bitcoin transactions.

We propose the SPV contract architecture that allows for managing the depth parameter $d$ -- the length of the \textit{confirmed} and \textit{pending} chain:

\begin{itemize}
    \item \textbf{Confirmed Chain}: the chain considered final with a high probability. This chain can be reverted, but 1) random reorg can happen only in very rare cases (reorganization of the Bitcoin blockchain on the \textit{n} last blocks); 2) the cost of the malicious reorg is very high.
    \item \textbf{Pending Chain}: consists of \textit{n} last blocks that are not yet confirmed and have a higher probability of being replaced due to potential reorganization.
\end{itemize}

There are different mechanisms for processing confirmed and pending chains:
\begin{itemize}
    \item Blocks of the confirmed chain represent the leaves of the Sparse Merkle Tree. Based on the root of SMT, it’s possible to prove that some transactions are included in the confirmed chain (it's very helpful for privacy \cite{spv-contract} and for the contract's storage compression).
    \item Blocks of the pending chain are stored in the form of a cache. It’s possible to have several alternative blocks for the same height in the pending chain; some will be orphaned when the mainchain is defined.
\end{itemize}

Any user can propose a new block header to the SPV contract. The proposed block header undergoes validation to ensure that it:
\begin{itemize}
    \item Has a valid structure
    \item References an existing previous block
    \item Satisfies the difficulty target
\end{itemize}
If all conditions are met, the block header is stored within the contract and becomes part of the mainchain if its total difficulty is the biggest.

As we mentioned, despite the high probability of blocks being confirmed, the possibility of chain reorganization persists. The SPV contract allows for updating the confirmed chain if a longer chain is presented, thereby ensuring consistency with the heaviest chain rule.

The SPV contract allows users to prove the existence of a transaction by providing $\pi_{\mathsf{SPV}}(\mathsf{TX})$:
\begin{itemize}
    \item The transaction $\mathsf{TX}$
    \item The block hash
    \item The Merkle branch leading to the root
\end{itemize}

Successful verification triggers predefined actions based on the provided proof. \\

\begin{tikzpicture}[node distance=1cm and 1cm]
  \tikzset{
    blockchain/.style={rectangle, draw=black, thick,
                       minimum width=1cm, minimum height=1cm,
                       text centered, fill =green!30}
  }
    \tikzstyle{arrow}        = [thick,->,>=stealth]
    \tikzstyle{dashedarrow}  = [thick,dashed,->,>=stealth]
    \tikzstyle{manual} = [thick]
    \tikzstyle{block} = [
            rectangle, rounded corners,
            minimum width=4cm, minimum height=1cm,
            text centered, draw=black, fill=blue!30
        ]

    \node[blockchain] (btc) {Bitcoin};
    \node at ($(btc.east)+(0.3,0)$) {:};
    \node[blockchain, right=of btc] (b0) {$b_0$};
    \node[blockchain, right=of b0] (b1) {$b_1$};
     \node[blockchain, right=of b1] (b2) {$b_2$};
    \node[right=of b2] (dots) {$\dots$};
    \node[blockchain, right=of dots] (bn) {$b_{n}$};
    
    \draw[manual](b0.east) -- (b1.west);
    \draw[manual](b1.east) -- (b2.west);
    \draw[manual](b2.east) -- ++(1cm ,0);
    \draw[manual](bn.west) -- ++(-1cm ,0);

    \draw[arrow](b0.south) -- ++(0, -0.5cm);
    \draw[arrow](b1.south) -- ++(0, -0.5cm);
    \draw[arrow](b2.south) -- ++(0, -0.5cm);
    \draw[arrow](bn.south) -- ++(0, -0.5cm);

     \node[rectangle,rounded corners, draw=black, thick, minimum height=1cm, text centered,  fit=(btc.south west)(bn.south east), anchor=north west]
    at ([yshift=-0.5cm]btc.south west) (relayers) {Relayers};

    \node[block, below=of relayers, yshift=-1cm]   (spv) {SPV Contract};
    \node[block, below=of spv,       yshift=-0.5cm] (app) {Application};

    \draw[arrow] (relayers.south) -- node[pos=0.3,right]{$h_0,\dots,h_n$} (spv.north);

    \draw[->, bend left, thick]  ([xshift=-0.5cm]app.north) to node[left]{\texttt{exists()}}  ([xshift=-0.5cm]spv.south);
    \draw[->, bend left, thick] ([xshift= 0.5cm]spv.south) to node[right]{\texttt{true}}    ([xshift= 0.5cm]app.north);

    \draw[arrow](app.east) -- ++(2cm, 0) node[midway, above] {\shortstack{\texttt{execute()}}};

    \node[block, left=of app, xshift=-1cm] (user){\faUser \; User};
    \draw[arrow](user.east)--node[above]{\texttt{call(),}} node[below]{$\pi_{\mathsf{SPV}}\mathsf{(TX)}$}(app.west);

    \coordinate (lineY) at ([yshift=-1cm]relayers.south);
    \draw[dashed]($(user.west|-lineY)$) -- ($(relayers.east|-lineY)$);
  
\end{tikzpicture}

\subsection{Compression of Bitcoin History with SNARKs}
When we investigated the rationality of launching the SPV node, we found that synchronizing the entire chain from the genesis block is extremely expensive. An internal implementation of the SPV contract, which can be found here \footnote[1]{\url{https://github.com/distributed-lab/spv-contracts/tree/master}}, showed that the verification of one block header consumes around 130k gas. As of June 16, 2025, the cost of SPV synchronization on the Ethereum mainnet is approximately 775,300 USD. 

We assume this cost can be reduced by replacing the verification of all blocks with a recursive proof up to a particular checkpoint (for example, having 900,000 blocks in total, we can prove 890,000 of them and then synchronize the remaining 10,000 as described above). In this case the contract initiator creates the proof $\pi_{\mathsf{hh}}$ for the following relation:
\begin{gather*}
        \mathcal{R}_{\mathsf{hh}} = \{\mathsf{header}^{n-1}; h_0, h_n: \\
        \forall i \in[1,n]:
        \\
        h_i \in \mathcal{H} \ \land \\
        \mathsf{H_{sha256}}(h_{i-1}) = h_i.\mathsf{prev} \ \land
        \\
        h_i.\mathsf{time} \geq \mathsf{med(h_{i-11}, \dots , h_{i-1})} \ \land \\
        \mathsf{H_{sha256}}(h_i)<\mathsf{target} \}
\end{gather*}
One project that attempts to build an SPV STARK-based prover can be found here \footnote[2]{\url{https://github.com/ZeroSync/ZeroSync/tree/main}}. Our implementation of the PLONK-based SPV prover is currently in progress.

\section{Payment Channels and Lightning Network}
Payment channels \cite{lightning} is a technology that allows counterparties to lock their funds and send instant payments within this channel without posting on-chain intermediate transactions. We will refer to the simplified payment channel construction to reuse it later for "loan channels" in Wrapless.

\subsection{Funding Transaction}\label{sec:funding_tx}

Assume we have two parties, Alice and Bob, who want to create a payment channel for sending funds between each other off-chain with $a\bitcoin$ and $b\bitcoin$ amounts respectively. For it, they cooperatively create what's so-called a \textbf{``funding transaction''} with the respective pair of keys:
\begin{gather*}
sk_a \xleftarrow{R} \mathbb{F}_p, \quad \mathsf{P_a} = sk_a G \\
sk_b \xleftarrow{R} \mathbb{F}_p, \quad \mathsf{P_b} = sk_b G
\end{gather*}
To secure funds that can only be unlocked through mutual cooperation, payment channels utilize multi-signatures. Depending on the version, this can be \texttt{P2WSH} or \texttt{P2TR} output. In case of \texttt{P2WSH} using ECDSA signatures and compressed public keys, the spending condition (\texttt{scriptPubKey}) and fulfilling witness (\texttt{scriptSig}) are:

\begin{scriptSig}
\begin{align*}
  & \elem{\sigma_a} \elem{\sigma_b} & \bscomment{Push signatures to stack}
\end{align*}
\end{scriptSig}

\begin{scriptPubKey}
\begin{align*}
  &\elem{2} & \bscomment{ 2 signatures are required } \\
  &\elem{\mathsf{P_a}} \elem{\mathsf{P_b}} & \bscomment{ Alice's and Bob's public keys  } \\
  &\elem{2} & \bscomment{ Two public keys are to stack } \\
  &\opcode{OP\_CHECKMULTISIG} & \bscomment{ Check keys and sigs from stack }
\end{align*}
\end{scriptPubKey}

In case of \texttt{P2TR} aggregated Schnorr signature and aggregated $X$-only public key are used:

\begin{equation}
    \sigma_{\text{agg}} = \sigma_A \oplus \sigma_B \,, \mathsf{P}_{\text{agg}} = \mathsf{P_A} \oplus \mathsf{P_B}
\end{equation}

Depending on protocol $\oplus$ is some secure aggregation operation, but the most common approach can be found in the Musig2~\cite{musig2} paper. The only thing we need to know, that $\sigma_{\text{agg}}$ is a valid signature only for key $\mathsf{P}_{\text{agg}}$, so we can check:

\begin{scriptSig}
\begin{align*}
  & \elem{\sigma_{\text{agg}}}
\end{align*}
\end{scriptSig}

\begin{scriptPubKey}
\begin{align*}
  \elem{\mathsf{P}_{\text{agg}}} \opcode{OP\_CHECKSIG}
\end{align*}
\end{scriptPubKey}

Later on, we would mention it as $\mathsf{multisig}$ without specification of the underlying scheme. So, finally, the ``funding transaction'' is a transaction with one multisignature output, fulfilling amount $a\bitcoin + b\bitcoin$ :
\begin{gather*}
  \mathsf{TX_{fund}}\{
  ...;
  \mathsf{(a\bitcoin + b\bitcoin, multisig(P_a, P_b))}
  \}
\end{gather*}

\subsection{Commitment Transactions}
Payments in channels are made through \textbf{``commitment transactions''}. Each of the parties creates an asymmetric transaction that spends a multi-signature output from the $\mathsf{TX_{fund}}$:
\begin{enumerate}
    \item local party spends their output after locktime, or it can be spent by counterparty if they sign it using the local party's revocation pubkey $\mathsf{P^{rev}}$
    \item the second output can be spent by the counterparty  immediately
\end{enumerate}Let initial amounts $a_0 = a$,  $b_0 = b$, $N$ --- number of blocks both of them agreed to lock funds for, $\mathsf{P_A^{{rev}_0}}$ and $\mathsf{P_B^{{rev}_0}}$ --- Alice's and Bob's commitment revocation pubkeys. The commitment transactions for Alice and Bob are:
\begin{gather*}
  \text{Alice's:} \ \mathsf{TX_{{comm}_0}^A}\{(\mathsf{TX_{fund}, 0, (\sigma_A, \sigma_B)}); (a_0 \bitcoin, \mathsf{(addr_A \land LT(}N\mathsf{)) \lor (P_B \land P_A^{{rev}_0})}), (b_0\bitcoin, \mathsf{addr_B})\}
\end{gather*}

\begin{tabular}{m{4cm}|m{4cm}|m{6cm}}
  \textbf{Inputs} & \textbf{Outputs} & \textbf{Output Script} \\
  \hline
 \texttt{OP\_0 <$\sigma_A$> <$\sigma_B$>} & $a_0\bitcoin$  & \texttt{OP\_IF <N> OP\_CHECKLOCKTIMEVERIFY OP\_DROP $\mathsf{P_A}$ OP\_CHECKSIG OP\_ELSE 2 $\mathsf{P_B}$, $\mathsf{P_A^{{rev}_0}}$ OP\_CHECKMULTISIG 2 OP\_ENDIF} \small{(Alice can spend this output after N blocks have passed \textbf{OR} Bob can spend this output if he signs a transaction using Alice’s revocation pubkey)} \\
  \hline
  & $b_0\bitcoin$ & \texttt{OP\_DUP OP\_HASH160 OP\_PUSHBYTES\_20 <addr\_B> OP\_EQUALVERIFY OP\_CHECKSIG} \small{(Bob can spend this output immediately)} \\
\end{tabular}

\begin{gather*}
  \text{Bob's:} \ \mathsf{TX_{{comm}_0}^B}\{(\mathsf{TX_{fund}, 0, (\sigma_A, \sigma_B)}); (b_0 \bitcoin, \mathsf{(addr_B \land LT(}N\mathsf{)) \lor (P_A \land P_B^{{rev}_0})}), (a_0\bitcoin, \mathsf{addr_A})\}
\end{gather*}

\begin{tabular}{m{4cm}|m{4cm}|m{6cm}}
  \textbf{Inputs} & \textbf{Outputs} & \textbf{Output Script} \\
  \hline
  \texttt{OP\_0 <$\sigma_A$> <$\sigma_B$>} & $b_0\bitcoin$  & \texttt{OP\_IF <N> OP\_CHECKLOCKTIMEVERIFY OP\_DROP $\mathsf{P_B}$ OP\_CHECKSIG OP\_ELSE 2 $\mathsf{P_A}$, $\mathsf{P_B^{{rev}_0}}$ OP\_CHECKMULTISIG 2 OP\_ENDIF} \small{(Bob can spend this output after $N$ blocks have passed \textbf{OR} Alice can spend this output if she signs a transaction using Bob's revocation pubkey)} \\
  \hline
  & $a_0\bitcoin$ & \texttt{OP\_DUP OP\_HASH160 OP\_PUSHBYTES\_20 <addr\_A> OP\_EQUALVERIFY OP\_CHECKSIG} \small{(Alice can spend this output immediately)} \\
\end{tabular}

Index 0 means that this pair of transactions (and revocation pubkeys) is the first. When the first commitment transaction is created and parties exchange signatures, the $\mathsf{TX_{fund}}$ can be sent, opening the channel.

After each payment, balances and revocation pubkeys are changed. For example, if Alice wants to send $p$ to Bob through this channel, she constructs a new pair of commitment transactions:
\begin{gather*}
  \mathsf{TX_{{comm}_1}^A}\{(\mathsf{TX_{fund}, 0, (\ldots, \ldots)}); (a_1 \bitcoin, \mathsf{(addr_A \land LT(}N\mathsf{)) \lor (P_B \land P_A^{{rev}_1})}), (b_1\bitcoin, \mathsf{addr_B})\} \\
  \mathsf{TX_{{comm}_1}^B}\{(\mathsf{TX_{fund}, 0, (\ldots, \ldots)}); (b_1 \bitcoin, \mathsf{(addr_B \land LT(}N\mathsf{)) \lor (P_A \land P_B^{{rev}_1})}), (a_1\bitcoin, \mathsf{addr_A})\}
\end{gather*}

where $a_1 = a_0 - p$ and $b_1 = b_0 + p$. Alice sends signature $\mathsf{\sigma_A(TX_{{comm}_1}^A)}$ through \texttt{commitment\_signed} message and Bob must respond with \texttt{revoke\_and\_ack} that includes \texttt{per\_commitment\_secret}, which is a secret that was used in the creation of $\mathsf{P_B^{{rev}_0}}$ and \texttt{next\_per\_commitment\_point} for next revocation pubkey $\mathsf{P_B^{{rev}_2}}$. Then Bob sends $\mathsf{\sigma_B(TX_{{comm}_1}^B)}$ and Alice reveals her revocation keys.

The revocation public key is created as follows:
\begin{gather*}
  \mathsf{P_A^{{rev}_i}} = \mathsf{Q^{rev} \cdot Sha256(Q^{rev} || C_A^{{rev}_i}) + C_A^{{rev}_i}\cdot Sha256(Q^{rev} || Q^{rev})}
\end{gather*}

where $\mathsf{Q^{rev}}$ --- revocation basepoint from \texttt{open\_channel} or \texttt{accept\_channel} messages, $\mathsf{C_A^{{rev}_i}}$ --- per commitment point from \texttt{open\_channel} and \texttt{accept\_channel} or \texttt{revoke\_ack} (depending on the number of the commitment tx and party).

\subsection{HTLC Payments}
While direct balance updates work for simple payments within a channel, the Lightning Network primarily uses \textbf{Hashed Timelock Contracts (HTLCs)} for payments, especially for routed payments across multiple channels. However, even in a direct channel between Alice and Bob, HTLCs are the standard mechanism.

An HTLC is a conditional payment output in a Bitcoin transaction. It allows funds to be locked such that they can be claimed under specific conditions involving a cryptographic hash and a time limit. Specifically, an HTLC output can be spent in one of two ways:
\begin{enumerate}
    \item \textbf{By the recipient (e.g., Bob):} If they can provide the preimage (secret) $R$ such that its hash $H = \mathsf{Hash}(R)$ matches the hash specified in the HTLC script, before a specified timeout (\texttt{cltv\_expiry}).
    \item \textbf{By the sender (e.g., Alice):} If the recipient fails to provide the preimage before the timeout expires.
\end{enumerate}

When Alice wants to send a payment to Bob using an HTLC, the process involves modifying the commitment transactions. Alice generates a secret preimage $R$ and computes its hash $H$. She then proposes adding an HTLC to Bob via an \texttt{update\_add\_htlc} message. This message contains the payment amount, the hash $H$, and the timeout \texttt{cltv\_expiry}.

If Bob accepts the HTLC, both parties update their commitment transactions to reflect this pending payment. The new commitment transactions will include an additional output representing the HTLC. For instance, Alice's updated commitment transaction ($\mathsf{TX_{comm'}^A}$) might look like this (simplified):

\begin{table}[H]
  \centering
  \begin{tabular}{m{4cm}|m{3cm}|m{7cm}}
    \textbf{Inputs} & \textbf{Outputs} & \textbf{Output Script} \\
    \hline
    $\mathsf{multisig(P_a, P_b)}$ & $a_0 - h \bitcoin$ & Alice can spend after $N$ blocks \textbf{OR} Bob can spend using Alice's revocation key $\mathsf{P_A^{rev'}}$. \\
    \cline{2-3}
                    & $b_0 \bitcoin$ & Bob can spend immediately. \\
    \cline{2-3}
                    & $h \bitcoin$ & Bob claims with preimage $R$ (where $\mathsf{Hash}(R)=H$) before \texttt{cltv\_expiry} \textbf{OR} Alice claims after timeout (\texttt{cltv\_expiry}).
  \end{tabular}
\end{table}

Similarly, Bob's commitment transaction ($\mathsf{TX_{comm'}^B}$) will have a corresponding HTLC output structure, viewed from his perspective (Received HTLC):

\begin{table}[H]
  \centering
  \begin{tabular}{m{4cm}|m{3cm}|m{7cm}}
    \textbf{Inputs} & \textbf{Outputs} & \textbf{Output Script} \\
    \hline
    multiSig ($\mathsf{P_a} + \mathsf{P_b}$) & $b_0 \bitcoin$ & Bob can spend this output after $N$ blocks have passed \textbf{OR} Alice can spend this output using Bob's revocation key $\mathsf{P_B^{rev'}}$. \\
    \cline{2-3}
                    & $a_0 - h \bitcoin$ & Alice can spend this output immediately. \\
    \cline{2-3}
                    & $h \bitcoin$ & Bob claims with preimage $R$ (where $\mathsf{Hash}(R)=H$) before \texttt{cltv\_expiry} \textbf{OR} Alice claims after timeout (\texttt{cltv\_expiry}). (HTLC Received by Bob) \\
  \end{tabular}
\end{table}

These new commitment transactions ($\mathsf{TX_{comm'}^A}$, $\mathsf{TX_{comm'}^B}$) are signed and exchanged, following a process similar to that for simple balance updates, which effectively adds the HTLC to the channel state.

To complete the payment, Bob reveals the preimage $R$ to Alice by sending an \texttt{update\_fulfill\_htlc} message before the timeout. Upon receiving $R$, Alice can verify that $\mathsf{Hash}(R) = H$. Knowing $R$ allows Bob to definitively claim the HTLC output if the commitment transaction were ever broadcast. More importantly, in the off-chain context, receiving the \texttt{update\_fulfill\_htlc} signals to Alice that the payment is successful. Both parties then update their commitment transactions again to remove the fulfilled HTLC and adjust their main balances accordingly, again using \texttt{commitment\_signed} and \texttt{revoke\_and\_ack} to finalize the state change.

If Bob fails to provide the preimage before the \texttt{cltv\_expiry}, Alice can send an \texttt{update\_fail\_htlc} message, and both parties update their commitment transactions to remove the expired HTLC, returning the funds to Alice's balance.

\subsection{Closing Channel}
Closing a payment channel can occur in two primary ways: cooperatively (mutual close) or unilaterally (force close).

\subsubsection{Mutual Close}
This is the preferred method where both parties agree to close the channel.
\begin{enumerate}
    \item Alice and Bob agree on the final channel balances, say $a_k \bitcoin$ for Alice and $b_k \bitcoin$ for Bob.
    \item They cooperatively construct a single \textbf{``closing transaction''} ($\mathsf{TX_{close}}$) that spends the funding transaction output ($\mathsf{TX_{fund}}$).
    \item This transaction has outputs directly paying $a_k \bitcoin$ to Alice's final address ($\mathsf{addr_A}$) and $b_k \bitcoin$ to Bob's final address ($\mathsf{addr_B}$), minus transaction fees.
    \begin{gather*}
      \mathsf{TX_{close}}\{(\mathsf{TX_{fund}, 0, (\sigma_A, \sigma_B)}); (a_k \bitcoin, \mathsf{addr_A}), (b_k \bitcoin, \mathsf{addr_B})\}
    \end{gather*}
    \item Both parties sign this transaction using the keys associated with the funding multi-signature ($\mathsf{P_a}, \mathsf{P_b}$).
    \item Once signed by both, the transaction is broadcast to the Bitcoin network. The funds become available to each party after the transaction is confirmed. This method is fast and efficient, as it involves no time locks.
\end{enumerate}

\subsubsection{Force Close}
If one party becomes unresponsive or uncooperative, the other party can unilaterally close the channel by broadcasting their latest valid commitment transaction.

Assume Alice decides to force close using her latest commitment transaction, $\mathsf{TX_{{comm}_k}^A}$. This transaction reflects the agreed-upon state $k$, including Alice's main balance, Bob's main balance, and any active HTLCs. She broadcasts this transaction to the Bitcoin network.

\begin{enumerate}
    \item Bob's direct output (\texttt{to\_remote}, representing his main balance $b_k \bitcoin$) can be spent by Bob immediately once $\mathsf{TX_{{comm}_k}^A}$ confirms.
    \item Alice's direct output (\texttt{to\_local}, representing her main balance $a_k \bitcoin$) is encumbered by a time lock (\texttt{to\_self\_delay}, denoted as $N$ blocks in the previous section). Alice can only spend these funds after the time lock expires.
    \item \textbf{Revocation:} This time lock ($N$) on Alice's \texttt{to\_local} output provides a window for Bob to contest the closure if Alice broadcast an outdated commitment transaction (e.g., $\mathsf{TX_{{comm}_j}^A}$ where $j < k$). If Bob possesses the revocation secret corresponding to state $j$ (which Alice would have revealed when moving to state $j+1$), Bob can use the revocation path within the \texttt{to\_local} script to immediately claim Alice's funds ($a_j \bitcoin$). This acts as a strong deterrent against attempting to cheat by broadcasting old states.
    \item \textbf{Handling In-flight HTLCs:} If there were active HTLCs at state $k$, $\mathsf{TX_{{comm}_k}^A}$ would contain corresponding HTLC outputs. These outputs are resolved on-chain via second-level transactions after $\mathsf{TX_{{comm}_k}^A}$ confirms:
        \begin{itemize}
            \item \textbf{HTLCs Offered by Alice:} An output exists for each HTLC Alice offered.
                \begin{itemize}
                    \item \textit{Success Case:} Bob can spend this output by creating a second-level transaction that reveals the preimage $R$ before the HTLC's timeout (\texttt{cltv\_expiry}).
                    \item \textit{Timeout Case:} Alice can spend this output via a second-level transaction after the \texttt{cltv\_expiry} has passed.
                \end{itemize}
            \item \textbf{HTLCs Received by Alice:} An output exists for each HTLC Alice received.
                 \begin{itemize}
                    \item \textit{Success Case:} Alice can spend this output via a second-level transaction that reveals the preimage $R$ (if she knows it) before the HTLC's timeout (\texttt{cltv\_expiry}).
                    \item \textit{Timeout Case:} Bob can spend this output via a second-level transaction after the \texttt{cltv\_expiry} has passed.
                \end{itemize}
            \item \textit{Revocation for HTLCs:} Importantly, similar to the \texttt{to\_local} output, these HTLC outputs also have a revocation path. If Alice broadcasts an outdated $\mathsf{TX_{comm}^A}$, Bob can use the corresponding revocation key to claim these HTLC outputs immediately via a second-level transaction, regardless of the preimage or timeouts.
        \end{itemize}
    \item If Alice broadcasts the correct, latest state $\mathsf{TX_{{comm}_k}^A}$, Bob cannot use the revocation path on any output. Alice can spend her \texttt{to\_local} output after the time lock $N$ expires. The HTLC outputs are resolved on-chain as described above (success or timeout path) via second-level transactions initiated by the party entitled to the funds under those conditions.
\end{enumerate}
    
Force closes are less desirable as they require waiting for time locks (both \texttt{to\_self\_delay} on the main output and potentially \texttt{cltv\_expiry} on HTLC outputs) and involve broadcasting the more complex commitment transaction. Resolving HTLCs on-chain requires additional second-level transactions, potentially leading to higher transaction fees and significantly slower fund recovery compared to a mutual close.

\section{Collateral Debt Position}
A Collateral Debt Position (CDP) is a financial construct that enables the usage of an asset as collateral to secure a loan for another asset. Typically, CDP is utilized for pledging volatile assets, such as Bitcoin and Ethereum, and securing a loan in stablecoins with a predictable interest rate. This allows for two major gains:
\begin{itemize}
    \item Getting a stablecoin loan without selling a volatile asset, the user keeps a long position with the borrower's collateral asset
    \item If the user does not sell a volatile asset, there are no taxes for income to pay from getting a loan
\end{itemize}

The significant risk for the user is the potential for borrower liquidation when the price of the collateral asset falls to the minimum borrower collateralization ratio, after which position liquidation occurs, and the user retains the loan but loses the collateral to the liquidator.

Let Alice be a borrower and Bob be a lender. Alice pledges BTC as collateral; Bob provides a stablecoin loan to Alice. If Alice fully repays the loan, she can withdraw the collateral. If the borrower's collateralization ratio falls below the minimum level, any third party appointed as a liquidator can repay the loan and seize the borrower's collateral for their own use.

\subsection{Loan Origination}
Borrower Collateralization Ratio (BCR) is calculated as:
\begin{equation*}
    \mathsf{BCR = \frac{BCAA * OP}{LA}}
\end{equation*}
where BCAA is the borrower's collateral asset amount, OP is the oracle price, and LA is the loan amount. The protocol defines a minimum BCR (MBCR) value after which CDP liquidation could be initiated. 

\begin{enumerate}
    \item The loan origination process begins with Alice locking collateral in the CDP position contract. Alice also makes a loan request for a specific loan amount. If the loan is not originated, Alice can claim the collateral asset back at her address.
    \item Alice initiates a loan origination transaction with the following rules applied:
    \begin{itemize}
        \item The maximum loan amount (MLA) is calculated as: $\mathsf{MLA = BCAA * OP / 
 MBCR}$
        \item An additional fee for loan origination might be applied
        \item An additional interest rate for the loan might be applied
    \end{itemize}
\end{enumerate}

If $\mathsf{LA < MLA}$, the borrower can withdraw part of the collateral asset until $\mathsf{LA = MLA}$.

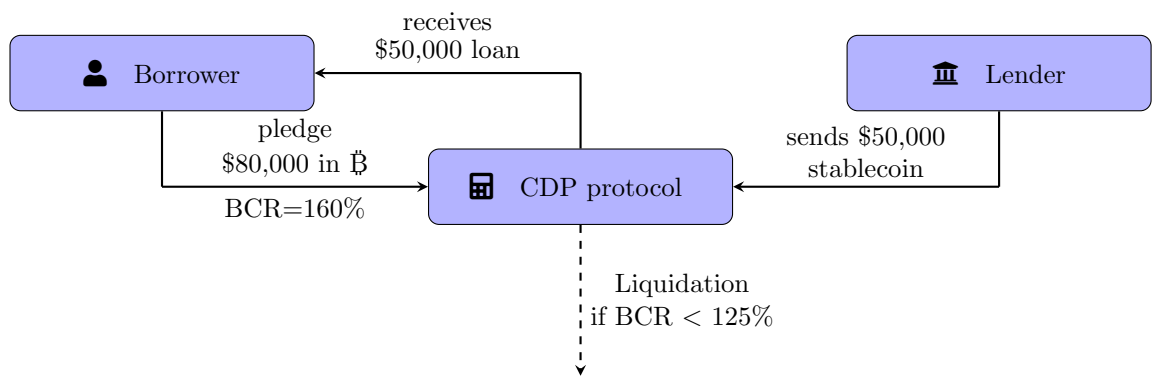
\begin{figure}[H]
\centering
\begin{tikzpicture}[node distance=1cm and 7cm, auto]
    % Styles
    \tikzstyle{block} = [rectangle, rounded corners, minimum width=4cm, minimum height=1cm, text centered, draw=black, fill=blue!30]
    \tikzstyle{arrow} = [thick,->,>=stealth]
    \tikzstyle{dashedarrow} = [thick,dashed,->,>=stealth]
    \tikzstyle{manual} = [thick]

    % Nodes
    \node (borrower) [block] {\faUser \quad Borrower};
    \node (lender) [block, right=of borrower] {\faUniversity \quad Lender};
    \node (cdp) [block, align=center, below=of $(borrower)!0.5!(lender)$] {
        \faCalculator \quad CDP protocol
    };

    \draw [manual] (lender.south) -- ++(0,-1cm);
    \draw [manual] (borrower.south) -- ++(0,-1cm);

    \draw [arrow] ($(lender.south)+(0,-1)$) -- ++(-3.5cm,0) node[midway, above] {\shortstack{sends \$50,000 \\ stablecoin}};
    \draw [manual] (cdp.north) -- ++(0,1cm);
    \draw [arrow] ($(borrower.east)+(3.5,0)$) -- ++(-3.5cm,0) node[midway, above] {\shortstack{receives \\ \$50,000 loan}};
    \draw [arrow] ($(borrower.south)+(0,-1)$) -- ++(3.5cm,0) node[midway, above] {\shortstack{pledge \\ \$80,000 in \bitcoin}} node[midway,below]{BCR=160\%};
    \draw[dashedarrow](cdp.south) -- ++ (0,-2cm)node[midway,right]{\shortstack{Liquidation \\ if BCR $<$ 125\%}};
    
\end{tikzpicture}
\caption{Loan origination via CDP}
\label{fig:cdp_loan_origination}
\end{figure}

\subsection{Borrower Liquidation Scenario}
In case the oracle price leads to $\mathsf{BCAA * OP / MBCR < OLA}$ (outstanding loan amount), the borrower liquidation process is triggered.

Bob or any other third party called a liquidator can initiate a transaction in which:
\begin{itemize}
    \item Liquidator repays the full amount of a loan in stablecoin plus accrued interest rate and fees (if any) outstanding
    \item Liquidator receives full borrower collateral asset
\end{itemize}

To avoid borrower liquidation, Alice should do either or both until BCR is restored to a minimal level equal to or higher than MBCR:
\begin{enumerate}
    \item Partially repay the loan
    \item Add borrower collateral
\end{enumerate}

\subsection{Loan Repayment and CDP Closure}
When Alice is ready to close the CDP position, the protocol requires:
\begin{enumerate}
    \item Alice repays the full amount of a loan in stablecoin plus accrued interest rate and fees (if any) to Bob
    \item Alice claims back the borrower's collateral asset
\end{enumerate}

Partial loan repayment is also possible. In this case, Alice can withdraw the borrower collateral until the $\mathsf{LA < MLA}$ based on the new collateral asset value.

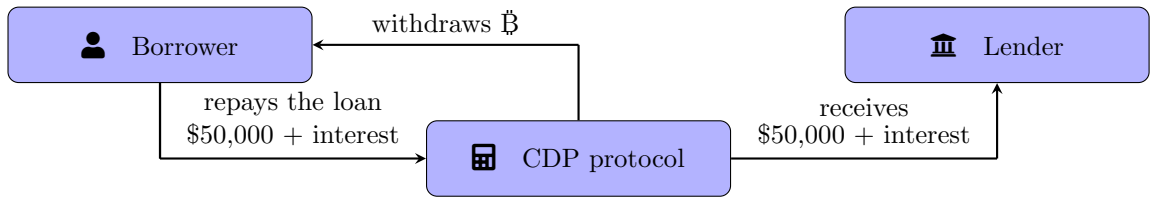
\begin{figure}[H]
\centering
\begin{tikzpicture}[node distance=1cm and 7cm, auto]
    % Styles
    \tikzstyle{block} = [rectangle, rounded corners, minimum width=4cm, minimum height=1cm, text centered, draw=black, fill=blue!30]
    \tikzstyle{arrow} = [thick,->,>=stealth]
    \tikzstyle{dashedarrow} = [thick,dashed,->,>=stealth]
    \tikzstyle{manual} = [thick]

    % Nodes
    \node (borrower) [block] {\faUser \quad Borrower};
    \node (lender) [block, right=of borrower] {\faUniversity \quad Lender};
    \node (cdp) [block, below=of $(borrower)!0.5!(lender)$] {
        \faCalculator \quad CDP protocol
    };

    \draw [arrow] ($(lender.south)+(0,-1)$) -- ++(0,1cm);
    \draw [manual] (borrower.south) -- ++(0,-1cm);
    \draw [manual] (cdp.north) -- ++(0,1cm);
    \draw [manual] ($(lender.south)+(0,-1)$) -- ++(-3.5cm,0) 
    node[midway, above] {\shortstack{receives \\ \$50,000 + interest}};
    \draw [manual] (borrower.south) -- ++(0,-1cm);
    \draw [arrow] ($(borrower.east)+(3.5,0)$) -- ++(-3.5cm,0) 
    node[midway, above] {\shortstack{withdraws \bitcoin}};
    \draw [arrow] ($(borrower.south)+(0,-1)$) -- ++(3.5cm,0) 
    node[midway, above] {\shortstack{repays the loan \\ \$50,000 + interest}};
   
\end{tikzpicture}
\caption{Loan repayment and CDP closure}
\label{fig:cdp_loan_closure}
\end{figure}

\chapter{Wrapless Protocol}

\section{Construction and properties} \label{sec:construction}
\textbf{Wrapless} is an analog of the CDP protocol that operates with native BTC. The primary features of the Wrapless protocol are trustlessness (through two-sided overcollateralization) and bidirectional liquidation. 
\begin{enumerate}
    \item \textbf{Trustlessness}. The parties create a cross-chain contract in a way that, if one of them manipulates, the other receives an economic advantage. The lender is protected by the collateral, the value of which exceeds the loan amount. The borrower is protected by the security deposit, which, together with the loan, exceeds the collateral value.
    \item \textbf{Bidirectional liquidation}. Existing CDP protocols are designed in a manner that assumes possible borrower liquidation when the price of the collateral asset falls to the minimum borrower collateralization ratio. However, at the same time, the borrower does not receive any advantage (except for decreasing the liquidation risk) if the collateral asset's value increases. The first version of Wrapless had a significant drawback that allowed the lender to seize the collateral at any point in time, particularly when its value increased substantially. We fixed this by creating the liquidation construction that works in both ways.
\end{enumerate}

Assume that Alice wants to lend USD from Bob on an EVM-based chain in exchange for her \bitcoin \ on the Bitcoin network. We can informally describe the protocol as follows:

\begin{description}
    \item[Agreement On Loan Details:] Bob and Alice must agree on the details of the future loan. It's possible to perform this procedure entirely off-chain or to create an order-based market in the form of a smart contract.
    
    \item[Loan Channel Establishment:] After details are agreed, Bob locks his $(b+c)$ USD on the loan contract and initiates the creation of a Lightning Network channel with Alice. Alice funds the channel with $a\bitcoin$ on Bob’s side and $0$ on Alice’s side. Alice broadcasts the funding transaction and provides SPV proof to the loan contract, marking the beginning of the lending process. Alice then receives $b\$$ on the EVM-based chain. If Alice doesn't submit a funding transaction to the Bitcoin network, Bob can return all deposited funds after the locktime $T_0$.
    
    \item[Loan Installment:] Alice sends $\frac{b(1+k)}{N}\$$ back to the loan contract to initiate an installment. In response, Bob must accept the installment and send $\frac{a}{N}\bitcoin$ to Alice through the loan channel. Both parties exchange signatures for their respective commitment transactions and submit them (excluding their signatures) to the loan contract. If both signatures are valid, the installment is considered finalized.
    
    \item[Loan Completion:] In case of a dispute (e.g., one party fails to submit the required signature for the commitment transaction either in the lightning channel or later to the loan contract) or upon successful repayment of all $N$ installments, the loan process concludes. The loan channel is closed, and the final commitment or closing transaction is submitted to the loan contract. Bob retains any remaining BTC; depending on how the loan channel was closed, one of the parties receives the security deposit.

    \item[Liquidation:] If the BTC price is changed and leads to the situation when one party is motivated to close the channel, an additional timeframe is opened. Depending on the direction in which the price moves, the borrower or liquidator must fund the channel or security deposit within this timeframe. If this funding has not been completed, the channel will be closed without penalty to the party responsible.
\end{description}

Below, we provide a more detailed explanation of the protocol flow, and from now on, Alice is the \textbf{borrower} $B$ and Bob is the \textbf{lender} $L$.

\section{Agreement On Loan Details} \label{sec:agreement}
Before the loan process begins, the lender and borrower should agree on the details. As we shortly mentioned, this can be done in on-chain and off-chain ways. Initially, let's define the parameters that need to be agreed upon between counterparties (later we will use $\mathcal{S}$ as a notation for the set of all parameters below):
\begin{itemize}
    \item [] $a$: collateral amount in \bitcoin
    \item [] $b$: loan amount in USD
    \item [] $k$: interest rate
    \item [] $c$: security deposit in USD provided by the lender. It can be unlocked by the borrower if their counterparty tries to cheat. If the entire lending process has been successful, the lender returns the security deposit.
    \item [] $N$: number of installments
    \item [] $T_0$: the time point before which the borrower must open the loan channel
    \item [] $T_1 \ldots T_N$: installment deadlines
    \item [] $\mathsf{lnid_B, lnid_L}$: lightning node IDs (the hash of $\mathsf{secp256k1}$ public key used for discovering and routing in Lightning Network)
    \item [] $\mathsf{P_B, P_L}$: the borrower's and lender's public keys
    \item [] $\mathsf{TX_{fund}}\{(\mathsf{TX_B, *, -});(\mathsf{a \bitcoin, multisig(P_B, P_L)})\}$: funding transaction for the loan channel (you can find it doesn't include signatures, they should be collected after forming the commitment transactions)
    \item [] $\mathsf{TX_{comm0}^B},\mathsf{TX_{comm0}^L}$: initial commitment transactions, formed as:
    \begin{gather*}
        \mathsf{TX_{{comm}_0}^B}\{(\mathsf{TX_{fund}, 0, (-, -)}); (0 \bitcoin, \mathsf{(addr_B \land LT_B) \lor (P_L \land P_B^{{rev}_0})}), (a\bitcoin, \mathsf{addr_L})\}
    \\
        \mathsf{TX_{{comm}_0}^L}\{(\mathsf{TX_{fund}, 0, (-, -)}); (a \bitcoin, \mathsf{(addr_L \land LT_L) \lor (P_B \land P_L^{{rev}_0})}), (0\bitcoin, \mathsf{addr_B})\}
\end{gather*}
    \item [] $\sigma_{L_b} \leftarrow \mathsf{sigGen_{ecdsa}(TX_{comm_0}^B, sk_L)}$ and $\sigma_{B_l} \leftarrow \mathsf{sigGen_{ecdsa}(TX_{comm_0}^L, sk_B)}$: signatures for initial commitment transactions
    \item [] $\mathcal{O}$: an oracle instance that returns the ratio between \bitcoin \ and USD and can initiate liquidation event. This oracle can be a service or a decentralized protocol, such as Uniswap \cite{uniswap}. Oracle cannot help one of the parties win the loan game we create, but rather signals a change in the price of collateral.
    \item [] $\mathsf{LR}_B$ and $\mathsf{LR}_L$: the liquidation ratio parameters
\end{itemize}

If all these details are agreed upon off-chain, the lender can initialize the contract by calling \texttt{createLoanOffer}(\(\mathcal{S}\)) method and providing the loan and security deposit (see Section \ref{sec:channel_establishment}). 

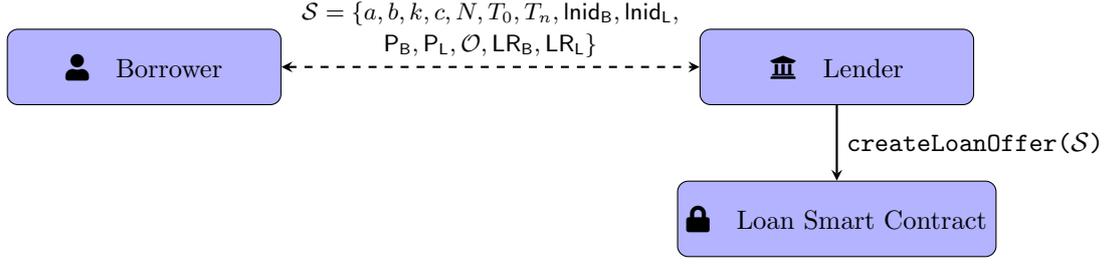
\begin{figure}[H]
    \centering
    \begin{tikzpicture}[node distance=2cm and 5.5cm, auto]
    % Styles
    \tikzstyle{block} = [rectangle, rounded corners, minimum width=3.6cm, minimum height=1cm, text centered, draw=black, fill=blue!30]
    \tikzstyle{arrow} = [thick,->,>=stealth]
    \tikzstyle{dashedarrow} = [thick,dashed,<->,>=stealth]
    \tikzstyle{parambox} = [rectangle, draw=black, fill=white, rounded corners, inner sep=6pt, minimum width=11.5cm]

    \node (borrower_top) [block] {\faUser \quad Borrower};
    \node (lender_top) [block, right=of borrower_top] {\faUniversity \quad Lender};

    \draw [dashedarrow] (borrower_top) -- node[above] {
    \small
    \shortstack{
        $\mathcal{S} = \{a, b, k, c, N, T_0, T_n, \mathsf{lnid_B}, \mathsf{lnid_L},$ \\
        $\mathsf{P_B}, \mathsf{P_L}, \mathcal{O}, \mathsf{LR_B}, \mathsf{LR_L}\}$
    }
} (lender_top);
    
    \node (escrow) [block, below=1cm of lender_top] {\faLock \quad Loan Smart Contract};

    \draw [arrow] (lender_top) -- node[right] {\texttt{createLoanOffer($\mathcal{S}$)}} (escrow);

\end{tikzpicture}
    \caption{Off-chain agreement}\label{fig:lender_accept_order_with_ order_icon}
\end{figure}

However, in practice, we can have a model where the order-matching mechanism exists on-chain; lenders initially define the loan conditions, and borrowers can select an option they are interested in:
\begin{enumerate}
    \item The lender defines the set of parameters $\mathcal{L}=\{min_a, max_a, \mathsf{BCR}, k, c, N, \Delta t, \mathsf{lnid_L}, \mathsf{P_L}, \mathcal{O}, \mathsf{LR}_B, \mathsf{LR}_L\}$, where $min_a, max_a$ - minimal and maximum acceptable collateral value, $\Delta t$ - installment period. The lender can offer various options $\mathcal{L}^*$ depending on the relationship between the parameters. 
    \item The borrower selects the option $\mathcal{L}_i$ from the list of available ones and provides their set of parameters by \texttt{requestLoan}(\(\mathcal{B}\)), $\mathcal{B}=\{a, T_0,\mathsf{lnid_B}, \mathsf{P_B},\mathsf{TX_{fund}, TX_{comm_0}^B}\}$
    \item If the lender agrees to provide the loan, they confirm it by calling \texttt{acceptLoan}(\(\mathsf{TX_{\!comm0}^L},\,\sigma_{L_b}\)). Additionally, the lender locks $(b+c)$ USD (details explained in Section \ref{sec:channel_establishment}).
    \item A final borrower's confirmation (method \texttt{finalize}(\(\sigma_{B_l},\,\pi_{\mathsf{SPV}}(\mathsf{TX}_{\mathsf{fund})}\))). Installment deadlines are calculated as $T_i = T_0 + i\cdot \Delta t, i \in [1, N]$. 
\end{enumerate}

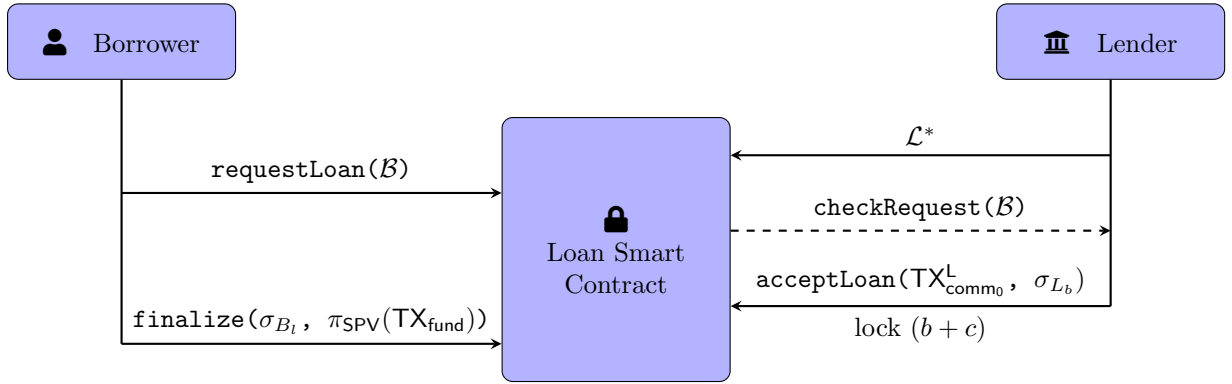
\begin{figure}[H]
    \centering
    \begin{tikzpicture}[node distance=1cm and 10cm, auto]
    % Styles
    \tikzstyle{block} = [rectangle, rounded corners, minimum width=3cm, minimum height=1cm, text centered, draw=black, fill=blue!30]
    \tikzstyle{arrow} = [thick,->,>=stealth]
    \tikzstyle{dashedarrow} = [thick,dashed,->,>=stealth]
    \tikzstyle{manual} = [thick]

    % Nodes
    \node (borrower) [block] {\faUser \quad Borrower};
    \node (lender) [block, right=of borrower] {\faUniversity \quad Lender};
    \node (escrow) [block, minimum height=3.5cm, text width=2cm, align=center, below=1cm of $(borrower)!0.5!(lender)$] {
        \faLock \\ Loan Smart Contract
    };

    \draw [manual] (lender.south) -- ++(0,-3cm);
    \draw [manual] (borrower.south) -- ++(0,-3.5cm);
    
    %Arrows
    \draw [arrow] ($(lender.south)+(0,-1)$) -- ++(-5cm,0) node[midway, above] {$\mathcal{L}^*$};
    \draw [arrow] ($(borrower.south)+(0,-1.5)$) -- ++(5cm,0) node[midway, above] {\texttt{requestLoan($\mathcal{B}$)}};
    \draw [dashedarrow] ($(escrow.east)+(0,0.25)$) -- ++(5cm,0) node[midway, above] {\texttt{checkRequest($\mathcal{B}$)}};
    \draw [arrow] ($(lender.south)+(0,-3)$) -- ++(-5cm,0) node[midway, above] {\texttt{acceptLoan($\mathsf{TX_{comm_0}^L}$, $\sigma_{L_b}$})} node[midway, below] {lock $(b+c)$};
    \draw [arrow] ($(borrower.south)+(0,-3.5)$) -- ++(5cm,0) node[midway, above] {\texttt{finalize($\sigma_{B_l}$, $\pi_{\mathsf{SPV}}(\mathsf{TX_{fund}})$)}} ;
    \end{tikzpicture}
    \caption{Borrower places order on Loan Smart Contract, creating an Order}
    \label{fig:borrower_order_creation}
\end{figure}
%update the diagram above with steps with on-chain agreements
%done?

Therefore, we can see that all parameters can be agreed upon in both ways: on-chain and off-chain. With the on-chain agreement process, it makes sense to wrap all functions with the minimal reasonable fee to protect each user from DoS of the counterparty.

\begin{ideablock}{On usage of the lender's public key}
    The lender's public key $\mathsf{P_L}$ is required for the borrower to verify the validity of $\mathsf{TX_{fund}}$ during channel establishment, as well as for the loan contract to verify its validity. Assume we have keys:

    \begin{gather*}
        sk_L \xleftarrow{R} \mathbb{F}_p, \quad \mathsf{P_L} = sk_L G \\
        sk_B \xleftarrow{R} \mathbb{F}_p, \quad \mathsf{P_B} = sk_B G
    \end{gather*}

    where $sk_L$ and $sk_B$ -- are the lender's and borrower's secret keys, respectively, that they set in the loan contract. So that borrower, after receiving the \texttt{open\_channel} \footnote[1]{\url{https://github.com/lightning/bolts/blob/ccfa38ed4f592c3711156bb4ded77f44ec01101d/02-peer-protocol.md\#the-open_channel-message}} message, can compare $\mathsf{P_L}$ to the received key in \texttt{funding\_pubkey} field of the message thus make sure that its the same lender, that accepted the order. For the loan contract, the importance comes from examining the provided funding transactions. Loan constructs \texttt{P2WSH} redeem script\footnote[2]{\url{https://github.com/bitcoin/bips/blob/9a56d3544eac1f949a747c251810f7a440d63fb9/bip-0141.mediawiki\#witness-program}}:

    \begin{redeemScript}
    \begin{align*}
        \opcode{OP\_2} \elem{\mathsf{P_L}} \elem{\mathsf{P_B}} \opcode{OP\_2} \opcode{OP\_CHECKMULTISIG}
    \end{align*}
    \end{redeemScript}

    hashes it $h = \text{Sha256}(\mathsf{redeemScript})$ and finally constructs \texttt{scriptPubKey} that it can compare to one in $\mathsf{TX_{fund}}$:

    \begin{scriptPubKey}
    \begin{align*}
        \opcode{OP\_0} \elem{h}
    \end{align*}
    \end{scriptPubKey}
\end{ideablock}

\section{Loan request acceptance and channel establishment} \label{sec:channel_establishment}
When all details are agreed, the lender locks $(b+c)$ USD on the loan contract. The loan $b$, can be taken by the borrower, but only if they open a loan channel on Bitcoin. A security deposit $c$ is being controlled by the loan contract.

\begin{ideablock}{Idea Block}
    There may be a concern that the lender needs to freeze the security deposit, $c$, before the loan is paid off. It creates an overcollateralization from the lender's side. However, 1) the lender can unlock the security deposit partially after covering the part of the loan by the borrower; 2) these funds can actually be used as liquidity for other protocols. Therefore, the deposit plays a crucial role in the security of the lending process, and at the same time, can serve as a source of additional income.
\end{ideablock}

Therefore, the single option for the borrower to take a loan is to send $\mathsf{TX_{fund}}$, which was previously agreed upon (the loan contract knows the specific Bitcoin transaction that must be sent). 
\begin{gather*}
    \mathsf{TX_{fund}}\{(\mathsf{TX_B, *, \sigma_B});(\mathsf{a \bitcoin, multisig(P_B, P_L)})\}
\end{gather*}

\begin{figure}[H]
    \centering
    \begin{tikzpicture}[node distance=2cm, auto]
        % Define styles for nodes and arrows
        \tikzstyle{block} = [rectangle, rounded corners, minimum width=3cm, minimum height=1cm, text centered, draw=black, fill=blue!30]
        \tikzstyle{arrow} = [thick,->,>=stealth]
        \tikzstyle{doublearrow} = [thick,<->,>=stealth]
        \tikzstyle{dashedarrow} = [thick,dashed,->,>=stealth]

        % Nodes
        \node (lender) [block] {\faUniversity \quad Lender};
        \node (channel) [block, right of=lender, xshift=4cm, fill=yellow!30] {\faBolt \quad Channel};
        \node (channeltext) [below of=channel, node distance=1.2cm] {$  \mathsf{TX_{fund}}\{*;\mathsf{(a\bitcoin, multisig(P_L, P_B))}, *\}$};
        \node (borrower) [block, right of=channel, xshift=4cm] {\faUser \quad Borrower};
        \node (escrow) [block, below of=channel, yshift=-1cm, xshift=3cm] {\faLock \quad Loan Smart Contract};

        % Arrows
        \draw [doublearrow] (lender.east) -- node[below, align=center] {} (channel.west);
        \draw [doublearrow] (channel.east) -- node[below, align=center] {} (borrower.west);
        \draw [arrow] (lender.south) |- node[pos=0.75, above] {\texttt{openChannel}$(\mathsf{TX_{fund}}, \pi_{\mathsf{SPV}}, \mathsf{LT_B, LT_L})$} (escrow.west);
    \end{tikzpicture}
    \caption{A loan channel creation}
    \label{fig:lender_borrower_channel}
\end{figure}
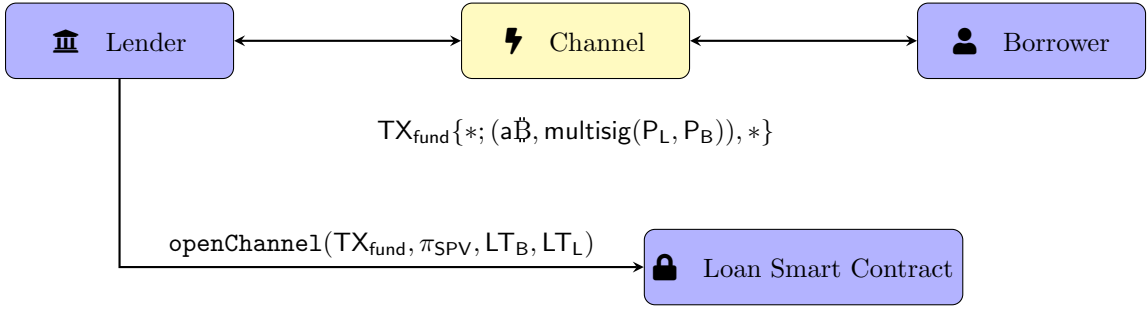

If the borrower sent $\mathsf{TX_{fund}}$, they can provide $\pi_{\mathsf{SPV}}(\mathsf{TX_{fund}})$ to the loan contract before $T_0$ (Figure~\ref{fig:lender_borrower_channel}). If the borrower does not open the channel (or fails to prove it before $T_0$), it means the lender can reclaim all $(b+c)$ USD (it's the reason why we proposed having a minimal, reasonable fee and an on-chain agreement for the loan). \texttt{openChannel}(\(\mathsf{TX}_{\mathsf{fund}},\,\pi_{\mathsf{SPV}})\) needs to be called to receive the loan.

The loan contract verifies the validity of $\pi_{\mathsf{SPV}}(\mathsf{TX_{fund}})$ by invoking the appropriate method of the SPV contract, which maintains a complete record of the Bitcoin blockchain history. If the proof is correct, the borrower can take a loan of $b$. Finally, $\mathsf{LT_B, LT_L}$ are locktimes for outputs to the local party for Alice and Bob in all commitment transactions, required for checks in further steps. 

\begin{comment}
\begin{figure}[H]
    \centering
    \begin{sequencediagram}
    \newthread{b}{\quad \quad \faUser \ Borrower \quad \quad}{}
    \newthread{esc}{\quad \quad \faLock \ Escrow\quad \quad}{}
    \newthread{l}{\quad \quad \faUniversity \ Lender \quad \quad}{}

    \begin{messcall}{b}{\texttt{placeOrder}}{esc}
    \end{messcall}
    
    \begin{call}{esc}{\faFile \ Order}{l}{\texttt{acceptOrder}}
    \end{call}

    \begin{sdblock}{\faBolt \ Lightning}{}
        \begin{messcall}{l}{\texttt{open\_channel}}{b}
        \end{messcall}

        \begin{messcall}{b}{\texttt{accept\_channel}}{l}
        \end{messcall}

        \begin{call}{b}{Publish $\mathsf{TX_{fund}}$}{b}{}
        \end{call}
    \end{sdblock}

    \begin{messcall}{b}{\texttt{registerChannnel}}{esc}
    \end{messcall}

    \end{sequencediagram}
    \caption{Sequence diagram of operations done by participants of the protocol.}\label{fig:flow_seqdiagram}
\end{figure}
\end{comment}

\subsection{Closing the Loan Channel Instantly}
After the channel is established, it can be closed at any point in time. It makes no sense for the borrower to initiate it because they will lose the collateral. However, a reasonable question is why the lender does not close the channel to receive the collateral, whose value exceeds the amount of the provided loan. 

We cannot influence the lender's decision to close the channel. Still, we can create conditions that make these attempts economically irrational --- specifically, by allowing the borrower to take a security deposit of $c$ in the case the lender closes the channel.

So, assume the channel was created and a loan was taken (no loan installments have been paid yet):
\begin{itemize}
    \item $\mathsf{TX_{fund}}$ was confirmed on-chain, $\mathsf{TX_{comm0}^B},\mathsf{TX_{comm0}^L}$ exist, each of them can be signed and propagated to the Bitcoin network
    \item The borrower has already received $b$, the lender controls $a$, $c$ is locked
\end{itemize}

There are several possible scenarios at this stage, so we define corresponding rules for them that the protocol must follow:

\begin{enumerate}
    \item The borrower closes the channel by co-signing and publishing the $\mathsf{TX_{comm_0}^B}$:
    \begin{gather*}
        \mathsf{TX_{{comm}_0}^B}\{(\mathsf{TX_{fund}, 0, (\sigma_{L_b}, \sigma_{B_b})}); (0 \bitcoin, \mathsf{(addr_B \land LT_B) \lor (P_L \land P_B^{{rev}_0})}), (a\bitcoin, \mathsf{addr_L})\}
    \end{gather*}
    \begin{enumerate}
        \item The borrower receives $0\bitcoin$
        \item The lender can take $a\bitcoin$ instantly
        \item The lender waits for $\mathsf{TX_{comm_0}^B}$ confirmation and provides $\pi_{\mathsf{SPV}}(\mathsf{TX_{comm_0}^B})$ to the loan contract (as a proof that the borrower initiated the channel close). This proof allows them to claim $c$ USD. 
    \end{enumerate}
    \item The lender closes the channel by co-signing and publishing the $\mathsf{TX_{comm_0}^L}$:
        \begin{gather*}
        \mathsf{TX_{{comm}_0}^L}\{(\mathsf{TX_{fund}, 0, (\sigma_{L_l}, \sigma_{B_l})}); (a \bitcoin, \mathsf{(addr_L \land LT_L) \lor (P_B \land P_L^{{rev}_0})}), (0\bitcoin, \mathsf{addr_B})\}
    \end{gather*}
    \begin{enumerate}
        \item The borrower receives $0\bitcoin$
        \item The lender can take $a\bitcoin$ after locktime $\mathsf{LT_L}$
        \item The borrower waits for $\mathsf{TX_{comm_0}^L}$ confirmation and provides $\pi_{\mathsf{SPV}}(\mathsf{TX_{comm_0}^L})$ to the loan contract (as a proof that the lender initiated the channel close). This proof allows them to take the security deposit. 
    \end{enumerate}
    \item Parties close the channel cooperatively using the transaction $\mathsf{TX_{close}}$:
        \begin{gather*}
            \mathsf{TX_{close}}\{(\mathsf{TX_{fund}}, 0, (\sigma_{L}, \sigma_{B}));(e\bitcoin, \mathsf{addr_L}),((a-e)\bitcoin, \mathsf{addr_B})\}
        \end{gather*}
    \begin{enumerate}
        \item The borrower receives $(a-e)\bitcoin$ instantly
        \item The lender receives $e\bitcoin$ instantly
        \item The lender waits for $\mathsf{TX_{close}}$ confirmation and provides $\pi_{\mathsf{SPV}}(\mathsf{TX_{close}})$ to the loan contract. By default, we presume that the lender can take the security deposit; however, it is possible to organize conditions under which the security deposit is distributed between counterparties based on their agreement.
    \end{enumerate}
    \item Parties continue working by paying loan installments and creating new commitment transactions in the channel.
\end{enumerate}

\begin{center}
\begin{tabular}{ |C{4em}|C{7em}|C{6em}|C{9em}|C{6em}| } 
\hline
\multirow{2}{*}{} & Borrower closes & Lender closes & Cooperative &  Normal flow \\

\hline
\arrayrulecolor{black}

Lender & \cellcolor{green!20} $(a+c)$ 
       & \cellcolor{red!20} $a$   
       & {\cellcolor{gray!20}up to agreement}
       & \cellcolor{green!20} $b\cdot(1+k)+c$  \\

\arrayrulecolor{black}
\cline{1-3}
\cline{5-5}
Borrower & \cellcolor{red!20} $b$
         & \cellcolor{green!20} $b+c$  
         & {\cellcolor{gray!20}}
         & \cellcolor{green!20} $a$ \\
\hline

\end{tabular}
\end{center}

\section{Loan Installment}

After a channel is established and registered on the loan smart contract, the borrower must follow the installment conditions, paying fractions of the loan until the corresponding deadlines, $T_1,..., T_N$. 

The payment $i\in[1,N]$ is made by the \texttt{payInstallment(...)} method, providing $\frac{b(1+k)}{N}$ USD to the contract. Additionally, the borrower needs to attach the $\mathsf{TX_{comm_i}^B}$ with the new BTC distribution in the channel:
\begin{gather*}
    \mathsf{TX^B_{{comm}_i}} = \{ 
    (\mathsf{TX_{fund}}, 0, (-, -));
    (\frac{a\cdot i}{N}\bitcoin, (\mathsf{addr_B} \land \mathsf{LT_B}) \lor (\mathsf{P_B^{{rev}_i}} \land \mathsf{P_L}), (a - \frac{a\cdot i}{N} \bitcoin, \mathsf{addr_L}))
    \}
\end{gather*}

This action forces the lender to accept the installment using the \texttt{takeInstallment(...)} method. This method requires providing the signature $\sigma_{L_{bi}}\leftarrow \mathsf{sigGen_{ecdsa}}(\mathsf{TX_{comm_i}^B, sk_L})$ and
the transaction $\mathsf{TX_{comm_i}^L}$, formed as:
\begin{gather*}
    \mathsf{TX^L_{{comm}_i}} = \{ 
    (\mathsf{TX_{fund}}, 0, (-, -));
    (a-\frac{a\cdot i}{N}\bitcoin, (\mathsf{addr_L} \land \mathsf{LT_L}) \lor (\mathsf{P_L^{{rev}_i}} \land \mathsf{P_B}), (\frac{a\cdot i}{N} \bitcoin, \mathsf{addr_B}))
    \}
\end{gather*}
The next step requires the borrower to provide the signature $\sigma_{B_{li}}\leftarrow \mathsf{sigGen_{ecdsa}}(\mathsf{TX_{comm_i}^L, sk_B})$ and the secret used for the previous $\mathsf{TX_{comm_{i-1}}^B}$. Finally, the lender must respond with the secret committed to $\mathsf{TX_{comm_{i-1}}^L}$. 

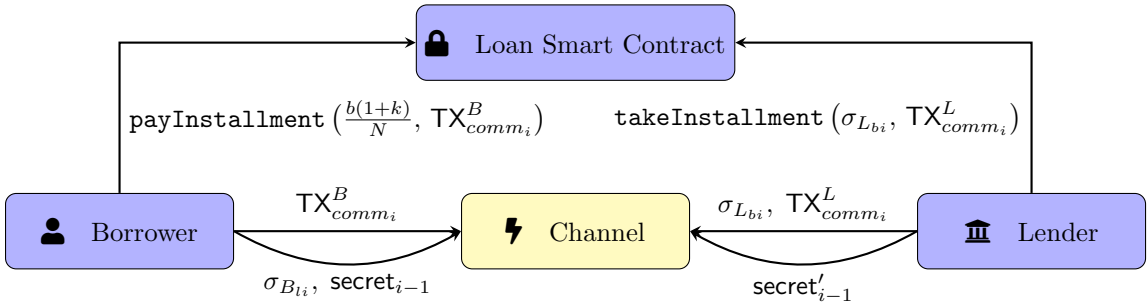
\begin{figure}[H]
    \centering
    \begin{tikzpicture}[node distance=2cm, auto]
        % Styles
        \tikzstyle{block} = [
            rectangle, rounded corners,
            minimum width=3cm, minimum height=1cm,
            text centered, draw=black, fill=blue!30
        ]
        \tikzstyle{channelblock} = [
            rectangle, rounded corners,
            minimum width=3cm, minimum height=1cm,
            text centered, draw=black, fill=yellow!30
        ]
        \tikzstyle{arrow} = [thick,->,>=stealth]

        % Nodes
         \node (borrower) [block] {\faUser \quad Borrower};
        \node (channel) [channelblock, right of=borrower, xshift=4cm] {\faBolt \quad Channel};
        \node (lender) [block, right of=channel, xshift=4cm] {\faUniversity \quad Lender};
        \node (escrow) [block, above of=channel, yshift=0.5cm] {\faLock \quad Loan Smart Contract};

        % On-chain calls
        \draw [arrow]
            (borrower.north) |- 
            node[pos=0.25, right]
                {\texttt{payInstallment}\,$\bigl(\tfrac{b(1+k)}{N},\,\mathsf{TX}^B_{comm_i}\bigr)$}
            (escrow.west);

        \draw [arrow]
            (lender.north) |-
            node[pos=0.25, left]
                {\texttt{takeInstallment}\,$\bigl(\sigma_{L_{bi}},\,\mathsf{TX}^L_{comm_i}\bigr)$}
            (escrow.east);

        % Off-chain commitment exchange (step 1 & 2)
        \draw [arrow]
            (borrower.east) -- 
            node[above, align=center] {$\mathsf{TX}^B_{comm_i}$}
            (channel.west);

        \draw [arrow]
            (lender.west) -- 
            node[above, align=center]
                {$\sigma_{L_{bi}},\;\mathsf{TX}^L_{comm_i}$}
            (channel.east);

        % Off-chain revocation exchange (step 3 & 4)
        \draw [arrow]
            (borrower.east) to[bend right=30]
            node[below, align=center]
                {$\sigma_{B_{li}},\;\mathsf{secret}_{i-1}$}
            (channel.west);

        \draw [arrow]
            (lender.west) to[bend left=30]
            node[below, align=center]
                {$\mathsf{secret}'_{i-1}$}
            (channel.east);
    \end{tikzpicture}
    \caption{Installment‐and‐commitment protocol: on-chain calls and off-chain handshake}
    \label{fig:payment_installment_with_channel_dashed}
\end{figure}

The last installment must lead to the creation of the $\mathsf{TX_{close}}$ Bitcoin transaction where the borrower takes all collateral from the channel.
\begin{gather*}
    \mathsf{TX_{close}}\{(\mathsf{TX_{fund}}, 0, (\sigma_{L}, \sigma_{B}));(0\bitcoin, \mathsf{addr_L}),(a\bitcoin, \mathsf{addr_B})\}
\end{gather*}

\subsection{Dispute Resolution}
The protocol incorporates mechanisms to handle scenarios where either party attempts to deviate from the agreed-upon rules or fails to fulfill their obligations.

\begin{enumerate}
    \item \textbf{Borrower fails to pay installment:} If the borrower does not call \texttt{payInstallment(...)} before the corresponding deadline $T_i$, the lender can initiate a channel closure procedure by broadcasting the latest, mutually signed commitment transaction $\mathsf{TX^L_{comm_{i-1}}}$ to the network. Then, the lender can take the security deposit -- the loan contract releases it if the borrower fails to make the installment payment. 

    \item \textbf{Lender attempts to publish outdated channel state:} The construction of the lightning channel protects against this case. If the lender attempts to close the channel with one of the previous commitment transactions, they provide the lender with the ability to withdraw all funds from the channel. We can also cover it with the ability to take the security deposit by the borrower; in this case, the borrower extracts the maximum value: $(a+b+c)$.

    \item \textbf{Lender fails to cooperate after installment payment:} If the borrower pays the installment via \texttt{payInstallment(...)}, but the lender doesn't accept it (doesn't provide the necessary signature ($\mathsf{\sigma_{L}(TX^B_{comm_i})}$) for the updated commitment transaction in the channel or fails to register it (\texttt{takeInstallment(...)}) with the loan contract) within a specified timeframe, the borrower can initiate a dispute. The dispute involves closing the channel with the latest state and retaining the security deposit by the borrower. 

    \item \textbf{Submitting incorrect data to loan contract:} If either party submits an invalid signature or an incorrectly constructed commitment transaction/revocation key, the loan contract's verification checks will fail and the transaction will be rejected. Such actions can be subject to penalties, such as forfeiture of the security deposit after a dispute period (like in the previous example).
    
\end{enumerate}

These mechanisms aim to ensure that rational actors are incentivized to follow the protocol honestly, as attempting to cheat typically results in a financial loss greater than any potential gain. The security deposit held in the loan contract plays a crucial role in penalizing malicious behavior and compensating the honest party.

\section{Liquidation}
The last and the most interesting part we want to cover is the case where the collateral price is changed. We can show initial loan agreement details schematically, in Figure \ref{fig:init}:
\begin{figure}[H]
    \centering
    \includegraphics[width=0.8\linewidth]{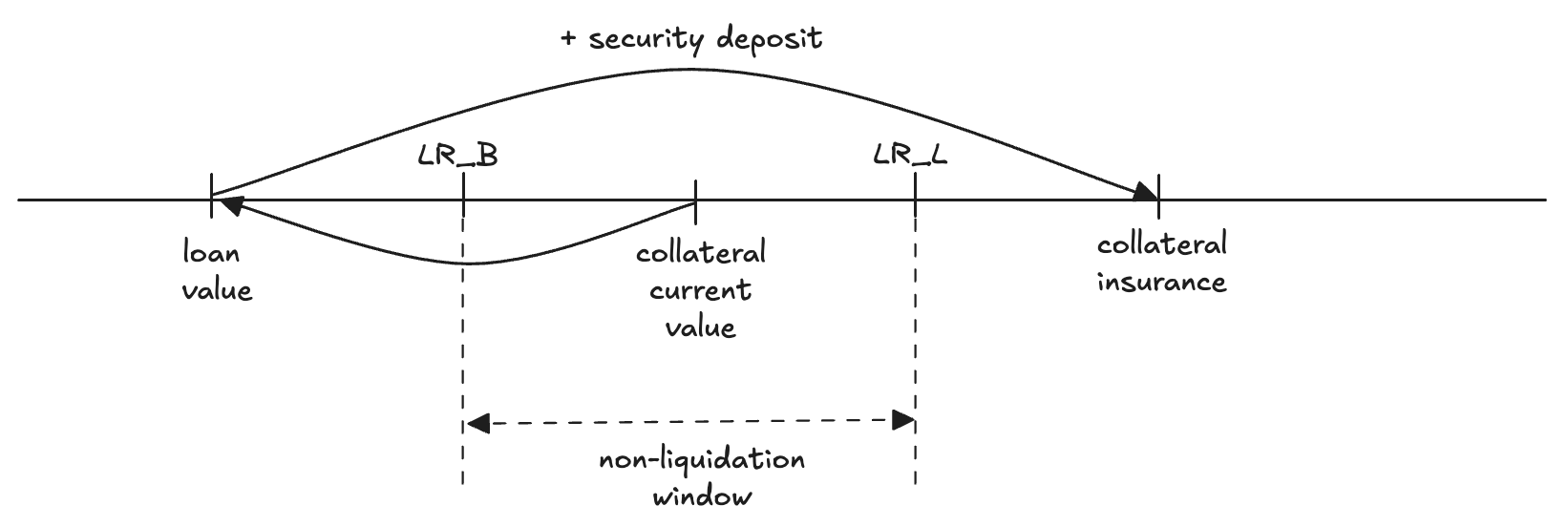}
    \caption{Initial loan parameters}
    \label{fig:init}
\end{figure}
We need to protect both parties in situations where changes in collateral prices can motivate the counterparty to act dishonestly:

\begin{enumerate}
    \item If the collateral price goes down and creates an undercollateralized loan, the borrower stops paying installments, and the lender loses the funds he loaned
    \item If the collateral price goes up and exceeds the loan and security deposit, the lender closes the channel, and the borrower loses the collateral
\end{enumerate}

Let's assume we have an oracle instance $\mathcal{O}$ integrated with the loan contract. This oracle provides the current exchange rate $r$ between USD and BTC on a regular basis (1\bitcoin \ = $r$ USD).

\subsection{Borrower's position liquidation}
If the oracle provides $r$, which is $r\cdot a = \mathsf{LR}_B$, it triggers the liquidation event because if the collateral prices continue going down, it will lead to an uncollateralized loan. 
\begin{figure}[H]
    \centering
    \includegraphics[width=0.8\linewidth]{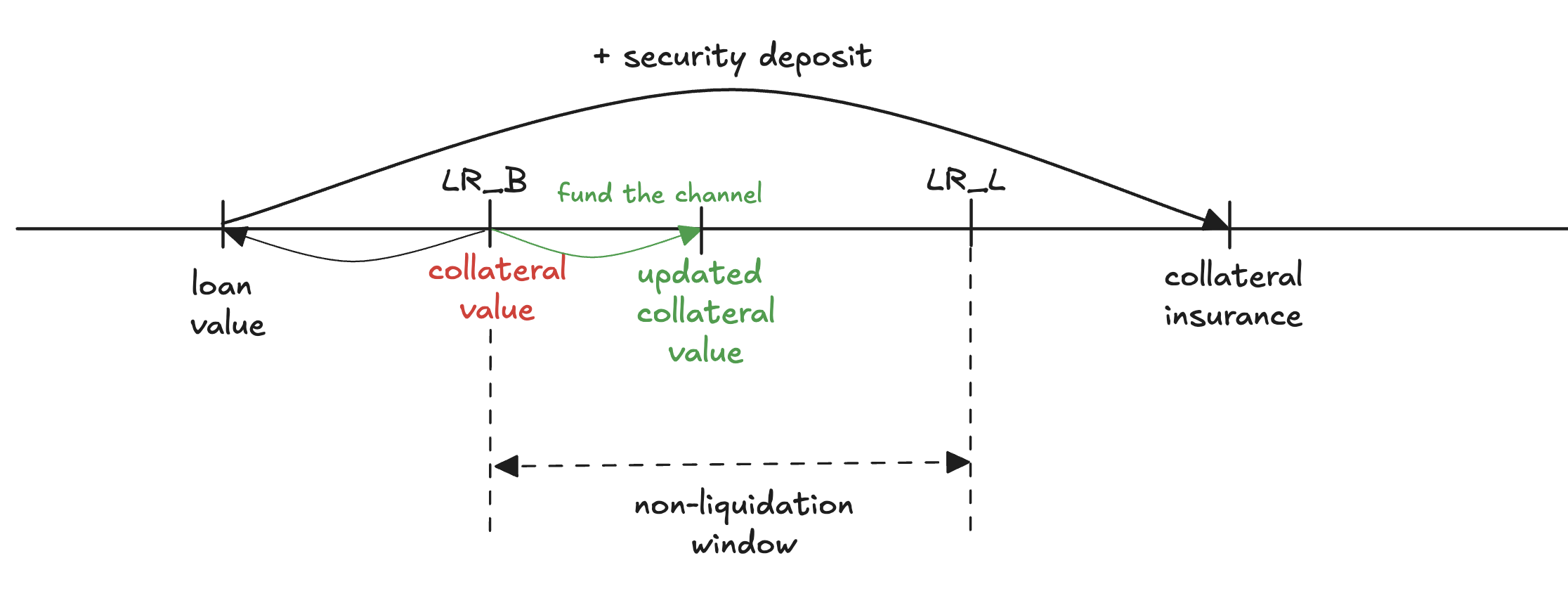}
    \caption{Funding the channel to cancel the liquidation}
    \label{fig:borrower-liquidation}
\end{figure}

The liquidation process consists of two stages. The first stage includes the timeframe during which the borrower can fund the channel and cancel the liquidation, thereby increasing the collateral value above $\mathsf{LR}_B$. 

If the borrower fails to fund the channel within the designated timeframe, the lender can close the channel without incurring a penalty. 

\subsection{Lender's position liquidation}
In this case, we assume the oracle provides $r$, such that $r\cdot a = \mathsf{LR}_L$. We need to initiate the liquidation event before the collateral price exceeds $(b+c)$, as this would render the lender economically motivated to close the channel.
\begin{figure} [H]
    \centering
    \includegraphics[width=0.8\linewidth]{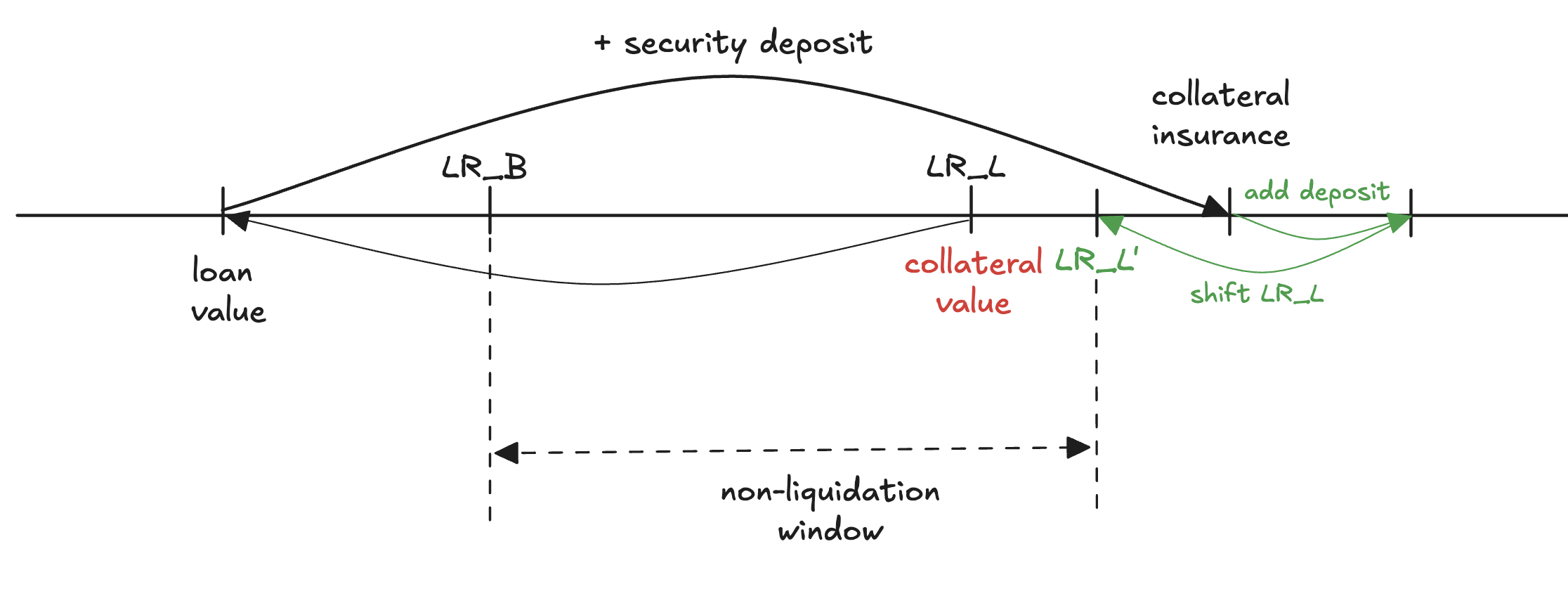}
    \caption{Funding the security deposit to cancel the liquidation}
    \label{fig:lender-liquidation}
\end{figure}

The liquidation procedure is also divided into two stages: 1 -- the period when the lender can increase the security deposit (moving $\mathsf{LR}_L$ value up), and 2 -- final liquidation if the lender didn't do that (in this case, the borrower can close the channel and take the security deposit).

\chapter{SPV contract specification}
\begin{itemize}
    \item[] \texttt{addBlockHeader}(\(block\_header\_raw\))
    \begin{itemize}
        \item[] Purpose: This function enables the addition of a new Bitcoin block header to the SPV contract's internal block header chain, allowing for simplified payment verification.
        \item[] Internal steps:
        \begin{enumerate}
            \item \textbf{Initial Parameter Parsing and Validation:} The function first parses the provided \texttt{blockHeaderRaw} input and performs several initial checks:
            \begin{itemize}[nosep]
                \item[-] ensuring that the length of \texttt{blockHeaderRaw} is exactly 80 bytes.
                \item[-] verifying that a block with the same hash has not been previously added to the contract's chain.
                \item[-] confirming that the \texttt{prevBlockHash} extracted from \texttt{blockHeaderRaw} is already present within the contract's block tree.
            \end{itemize}
            \item \textbf{Block Metadata Determination:} Following initial validation, the contract determines essential metadata for the new block, including its \texttt{height}, \texttt{target}, and \texttt{medianTime}.
            \item \textbf{Block Content Validation:} A validation of the block's content is then performed. This includes:
            \begin{itemize}[nosep]
                \item[-] verifying that the block's \texttt{target} (difficulty) aligns with the network's current expected \texttt{target}.
                \item[-] ensuring that \(\texttt{blockHash} \le \texttt{blockTarget}\), confirming it meets the proof-of-work requirement.
                \item[-] checking that the \(\texttt{blockTime} \ge \texttt{medianTime}\), preventing timestamp manipulation.
              \end{itemize}
            \item \textbf{Block Addition to Chain:} Upon successful validation, the block is added to the contract's internal block header chain. This process can result in one of three outcomes: the new block extends the main chain, becomes an alternative (fork) block, or triggers a reorganization of the main chain.
            \item \textbf{Event Emission:} A \texttt{BlockHeaderAdded}(block\_height, block\_hash) event is emitted.
        \end{enumerate}
    \end{itemize}
    %-----------------------------
    \item[] \texttt{addBlockHeaderBatch}(\(block\_header\_raw\_arr\))
    \begin{itemize}
        \item[] Purpose: This function facilitates the addition of a batch (array) of Bitcoin block headers to the SPV contract's internal chain, enabling more efficient synchronization.
        \item[] Internal steps:
        \begin{enumerate}
            \item \textbf{Batch Processing and Initial Validation:} The function iterates through each \texttt{blockHeaderRaw} element within the \texttt{blockHeaderRawArr}. For each individual block header, it performs initial parsing and validations analogous to those in \texttt{addBlockHeader}:
            \begin{itemize}[nosep]
                \item[-] ensuring that the length of \texttt{blockHeaderRaw} is exactly 80 bytes.
                \item[-] verifying that a block with the same hash has not been previously added.
                \item[-] confirming that the \texttt{prevBlockHash} extracted from \texttt{blockHeaderRaw} is already present within the contract's block tree.
            \end{itemize}
            \item \textbf{Block Metadata Determination:} For each block, its \texttt{height}, \texttt{target}, and \texttt{medianTime} are determined.
            \item \textbf{Block Content Validation:} For each block, a comprehensive validation of its content is performed, identical to the checks in \texttt{addBlockHeader} with the next optimizations:
            \begin{itemize}[nosep]
                \item[-] For subsequent blocks within the \texttt{blockHeaderRawArr}, the \texttt{prevBlockHash} can be directly validated against the hash of the immediately preceding block within the same input array, reducing the need for storage reads.
                \item[-] If the batch contains more than 11 block headers, the \texttt{medianTime} for blocks from the 12th onwards can be computed using the timestamps of the preceding blocks within the provided \texttt{blockHeaderRawArr}, rather than requiring reads from the contract's storage.
            \end{itemize}
            \item \textbf{Batch Block Addition to Chain:} Upon successful validation, each block in the array is added to the contract's internal block header chain. Similar to single block additions, this may extend the main chain, create alternative chains, or trigger reorganizations.
            \item \textbf{Event Emission:} For each successfully added block header, a \texttt{BlockHeaderAdded} \\(block\_height, block\_hash) event is emitted.
        \end{enumerate}
    \end{itemize}
     %-----------------------------
     \item[] \texttt{validateBlockHash}(\(block\_hash\))
     \begin{itemize}
         \item[] Purpose: This view function provides information about a given Bitcoin block header, specifically whether it resides on the SPV contract's recognized main chain and its number of confirmations.
         \item[] Parameters:
        \begin{itemize}[nosep]
            \item[-] \texttt{blockHash}: The hash of the Bitcoin block header to query.
        \end{itemize}
        \item[] Returns:
        \begin{itemize}[nosep]
            \item[-] (bool is\_in\_main\_chain, uint256 confirmations): A boolean indicating if the block is part of the main chain, and an unsigned integer representing its confirmation count.
        \end{itemize}
        \item[] Internal steps:
        \begin{enumerate}
            \item \textbf{Block Height Retrieval:} The function first retrieves the height of the block corresponding to the provided \texttt{blockHash} from the contract's storage.
            \item \textbf{Main Chain Check:} It then verifies if the provided \texttt{blockHash} is indeed part of the SPV contract's currently recognized main chain.
            \item \textbf{Confirmation Count Calculation:} The number of confirmations is determined by calculating the difference between the current height of the main chain and the height of the queried block. If \texttt{ blockHash} is not found in the main chain, its confirmation count is set to zero (0).
            \item \textbf{Value Return:} Finally, the function returns the determined is\_in\_main\_chain status and the calculated confirmations count.
        \end{enumerate}
     \end{itemize}
     %-----------------------------
     \item[] \texttt{verifyTx}(\(block\_hash, TX, \pi_{\mathsf{SPV}}(TX)\))
     \begin{itemize}
        \item[] Purpose: This view function verifies the inclusion of a specific Bitcoin \(TX\) within a \texttt{blockHash} using a provided \(\pi_{\mathsf{SPV}}(TX)\).
        \item[] Parameters:
        \begin{itemize}[nosep]
            \item \texttt{blockHash}: The hash of the Bitcoin block where the transaction is expected to be included.
            \item \(TX\): The raw bytes of the Bitcoin transaction.
            \item \(\pi_{\mathsf{SPV}}(TX)\): The Merkle proof demonstrating the transaction's inclusion in the block.
        \end{itemize}
        \item[] Returns:
        \begin{itemize}[nosep]
            \item bool: true if the transaction is successfully verified as part of the block's Merkle tree, false otherwise.
        \end{itemize}  
        \item[] Internal Steps:
        \begin{enumerate}
            \item \textbf{MR Calculation:} The function first computes a \texttt{calculatedMerkleRoot} by processing the provided \(TX\) and \(\pi_{\mathsf{SPV}}(TX)\).
            \item \textbf{Block MR Retrieval:} It then retrieves the canonical \texttt{blockMerkleRoot} associated with the given \texttt{blockHash} from the contract's stored block headers.
            \item \textbf{Return:} Finally, the function compares the \texttt{calculatedMerkleRoot} with the \\ \texttt{blockMerkleRoot}. The boolean result of this comparison is returned, indicating whether the transaction's inclusion in the block is valid.
        \end{enumerate}
     \end{itemize}
\end{itemize}

\chapter{Loan contract specification}

\begin{description}[style=nextline,leftmargin=2cm,labelwidth=0.5cm,labelsep=1cm]

  \item[\texttt{createLoanOffer}(\(\mathcal{S}\))]~\\[0.3em] 
  Purpose: This function facilitates the creation of a new loan offer by a \textbf{lender}.

  \medskip
  Internal Steps:
  \begin{enumerate}[label=\arabic*., left=1em]
    \item \textbf{Parameter Validation:} The function validates all input parameters, specifically:
      \begin{itemize}[nosep]
        \item ensuring that \(a\_min \le a\_max\).
        \item confirming the correct order of liquidation ratios: \(LR_B \le CR \le LR_L\).
        \item verifying that periods \(N, IP, IRP\) are all strictly positive values.
        \item ensuring that the lender's associated key \(P_L\) and node ID \(\mathsf{lnid}_L\) are non-zero.
        \item validating the installment response period (\(IRP\)) by checking that \(4 \times IRP \le IP\).
      \end{itemize}
    \item \textbf{State Persistence:} The function initializes and stores the following key parameters within the contract's storage, forming the immutable state of the loan offer:
      \[
        \begin{aligned}
          \texttt{min\_collateral\_amount}          &= a\_min,          &
          \texttt{security\_deposit}                &= c,               \\
          \texttt{max\_collateral\_amount}          &= a\_max,          &
          \texttt{interest\_rate}                   &= k,               \\
          \texttt{collateralization\_ratio}         &= CR,              &
          \texttt{installments\_count}              &= N,               \\
          \texttt{installment\_period}              &= IP,              &
          \texttt{response\_penalty}                &= RP,              \\
          \texttt{installment\_response\_period}    &= IRP,             &
          \texttt{lender\_key}                      &= P_L,             \\
          \texttt{first\_installment\_deadline}     &= T_0,             &
          \texttt{lnid\_l}                          &= \mathsf{lnid}_L, \\
          \texttt{oracle}                           &= \mathcal{O},     &
          \texttt{lr\_b}                            &= \mathsf{LR}_B,   \\
          \texttt{lr\_l}                            &= \mathsf{LR}_L.   
        \end{aligned}
      \]
    \item \textbf{Lender Penalty Deposit:} The \textbf{lender} is required to transfer an amount of \textbf{USD}, equivalent to the specified \texttt{responsePenalty}, to the contract. This transfer serves as a commitment associated with the offer.
    \item \textbf{Event Emission:} An \texttt{LoanOfferCreated}(offer\_id) event is emitted.
  \end{enumerate}

  \bigskip
  \item[\texttt{requestLoan}(\(\mathcal{B}\))]~\\[0.3em] 
  Purpose: This function allows a \textbf{borrower} to formally request a loan against an existing, available loan offer.

  \medskip
  Internal Steps:
  \begin{enumerate}[label=\arabic*., left=1em]
    \item \textbf{Parameter Validation:} The function first validates all provided input parameters. This includes:
    \begin{itemize}[nosep]
        \item ensuring that the Node ID \(\mathsf{lnid}_B\) and public key or address \(P_B\) are non-zero.
        \item confirming that the loan offer identified by \texttt{positionId} currently holds an \textit{Offered} status.
        \item verifying that the \(a\_min <= a <= a\_max\) established by the offer.
        \item checking that the remaining time until the \(T_0\) is at least \(2 \times IRP\).
    \end{itemize}
  
    \item \textbf{Offer State Update:} Upon successful validation, the internal state of the loan offer associated with \texttt{positionId} is updated. This involves changing its status to \textit{Requested} and incorporating relevant borrower details.
    \item \textbf{Borrower Penalty Deposit:} The \textbf{borrower} is required to transfer an amount of \textbf{USD} tokens, equivalent to the specified \(RP\), to the contract. This transfer serves as a commitment associated with the loan request.
    \item \textbf{Event Emission:} A \texttt{LoanRequested}(offer\_id) event is emitted.
  \end{enumerate}

  \bigskip
  \item[\texttt{checkRequest}(\(\mathcal{B}\))]~\\[0.3em] 
  Purpose: view-only validation of borrower’s submitted parameters.

  \medskip
  Internal Steps:
  \begin{enumerate}[label=\arabic*., left=1em]
    \item verify format and scripts of \texttt{TX\_fund} and \texttt{TX\_comm0\^B}
    \item check consistency with on-chain \(\{a,b,c,k\}\)
  \end{enumerate}

  \bigskip
  \item[\texttt{acceptLoan}(\(\mathsf{offer\_id},\mathsf{TX_{\!comm0}^L},\,\sigma_{L_b}\))]~\\[0.3em] 
  Purpose: This function allows the \textbf{lender} to accept a loan request, thereby finalizing the loan offer and proceeding with the fund transfer.

  \medskip
  Internal Steps:
  \begin{enumerate}[label=\arabic*., left=1em]
    \item \textbf{Parameter Validation:} The function first validates all provided input parameters. This includes:
    \begin{itemize}[nosep]
        \item ensuring that the loan offer identified by \texttt{positionId} currently holds a \textit{Requested} status.
        \item confirming that the transaction sender (\texttt{msg.sender}) is indeed the original creator of the loan offer (i.e., the lender associated with positionId).
        \item checking that the remaining time until the \(T_0\) is at least one \(IRP\).
        \item verify \(\sigma_{L_b}\) over \(\mathsf{TX_{\!comm0}^B}\) recovers \(P_L\).
    \end{itemize}
    \item \textbf{Offer State Update:} Upon successful validation, the provided \(\sigma_{L_b}\) and \(\mathsf{TX_{\!comm0}^B}\) are stored within the offer's state. Subsequently, the status of the loan offer associated with \texttt{positionId} is updated to "Accepted".
    \item \textbf{Loan Amount Calculation and Transfer:} The actual \(b\) is calculated as follows: \(b = \frac{a \times BTC\_price}{CR}\).
    The \(b + c\) is then transferred from the lender to the contract.
    \item \textbf{Event Emission:} A \texttt{LoanAccepted}(offer\_id) event is emitted.
  \end{enumerate}

  \bigskip
  \item[\texttt{openChannel}(\(\mathsf{offer\_id},\sigma_{B_l},\,\pi_{\mathsf{SPV}}TX\_fund)\))]~\\[0.3em] 
  Purpose: This function enables the \textbf{borrower} to finalize the loan offer, confirming the on-chain collateralization.

  \medskip
  Internal Steps:
  \begin{enumerate}[label=\arabic*., left=1em]
    \item \textbf{Parameter Validation:} The function first validates all provided input parameters. This includes:
    \begin{itemize}[nosep]
        \item ensuring that the loan offer identified by \texttt{positionId} currently holds an \textit{Accepted} status.
        \item confirming that the transaction sender (\texttt{msg.sender}) is indeed the \textbf{borrower} associated with positionId.
        \item verifying that this function call occurs before the \(T_0\).
        \item utilizing an SPV (Simplified Payment Verification) contract to verify \(\pi_{\mathsf{SPV}}(TX\_fund)\), confirming that the required collateral transaction has been successfully recorded on the BTC network.
    \end{itemize}
    \item \textbf{Offer State Update:} Upon successful validation, the provided \(\sigma_{B_l}\) is stored within the offer's state. Subsequently, the status of the loan offer associated with \texttt{positionId} is updated to \textit{Opened}.
    \item \textbf{Fund Transfer:} The contract transfers the total amount \(b + RP\) to the borrower's address.
    \item \textbf{Event Emission:} A \texttt{ChannelOpened}(offer\_id) event is emitted.
  \end{enumerate}

  \bigskip
  \item[\texttt{payInstallment}(\(\mathsf{offer\_id},\,TX^B\_{comm_i}\))]~\\[0.3em] 
  Purpose: This function enables the borrower to make a scheduled repayment for an active loan installment.

  \medskip
  Internal Steps:
  \begin{enumerate}[label=\arabic*., left=1em]
    \item \textbf{Parameter Validation:} The function first validates all provided input parameters. This includes:
    \begin{itemize}[nosep]
        \item ensuring that the loan offer identified by \texttt{positionId} currently holds an \textit{Opened} status.
        \item confirming that the transaction sender (\texttt{msg.sender}) is indeed the \textbf{borrower} associated with positionId.
        \item verifying that there are no outstanding or unfinished previous installments for this loan.
        \item checking that the remaining time until the deadline of the current installment is at least \(3 \times IRP\).
    \end{itemize}
    \item \textbf{Installment State Update:} The function marks the current installment as paid and stores the provided \(TX^B\_{comm_i}\) within the offer's state. Subsequently, the status of the current installment is updated to \textit{PaidInstallment}.
    \item \textbf{Installment Amount Transfer:} The borrower is required to transfer the calculated \texttt{installmentAmount} to the contract. The \texttt{installmentAmount} is determined by the following formula:
    \[
    \texttt{installmentAmount} = \frac{b}{N} + (b \times k)
    \]
    \item \textbf{Event Emission:} An \texttt{InstallmentPaid}(offer\_id, installment\_id) event is emitted.
  \end{enumerate}

  \bigskip
  \item[\texttt{takeInstallment}(\(\mathsf{offer\_id},\sigma_{L_{bi}},\,TX^L\_{comm_i}\))]~\\[0.3em] 
  Purpose: This function allows the \textbf{lender} to claim the principal and interest of a successfully paid installment.

  \medskip
  Internal Steps:
  \begin{enumerate}[label=\arabic*., left=1em]
    \item \textbf{Parameter Validation:} The function first validates all provided input parameters. This includes:
    \begin{itemize}[nosep]
        \item ensuring that the loan offer identified by \texttt{positionId} currently holds an \textit{Opened} status.
        \item confirming that the \texttt{msg.sender} is indeed the \textbf{lender} associated with positionId.
        \item verifying that the status of the current installment is \textit{PaidInstallment}.
        \item checking that the remaining time until the current installment's deadline is at least \(2 \times IRP\).
    \end{itemize}

    \item \textbf{Installment State Update:} The function updates the status of the current installment to \textit{TookInstallment} and stores the provided \(\sigma_{L_{bi}}\) and \(TX^L\_{comm_i}\) within the offer's state.
    \item \textbf{Event Emission:} An \texttt{InstallmentAccepted}(offer\_id, installment\_id) event is emitted.
  \end{enumerate}

  \bigskip
  \item[\texttt{revealRevocationKeyBorrower}(\(\mathsf{offer\_id}, \sigma_{B_l},PRK_b\))]~\\[0.3em] 
  Purpose: This function enables the \textbf{borrower} to disclose their borrowerPrevRevocationKey (\(PRK_b\)) for the current installment, an essential step in the loan's lifecycle.

  \medskip
  Internal Steps:
  \begin{enumerate}[label=\arabic*., left=1em]
    \item \textbf{Parameter Validation:} The function first validates all provided input parameters. This includes:
    \begin{itemize}[nosep]
        \item ensuring that the loan offer identified by \texttt{positionId} currently holds an \textit{Opened} status.
        \item confirming that the \texttt{msg.sender} is indeed the \textbf{borrower} associated with positionId.
        \item verifying that the status of the current installment is \textit{TookInstallment}.
        \item checking that the remaining time until the current installment's deadline is at least \(IRP\).
    \end{itemize}

    \item \textbf{Installment State Update:} The function updates the status of the current installment to \textit{BorrowerRevocationKey} and stores the provided \(\sigma_{B_l}\) and \(PRK_b\) within the offer's state.
    \item \textbf{Event Emission:} A \texttt{BorrowerRevocationKeyRevealed}(offer\_id, installment\_id) event is emitted.
  \end{enumerate}

  \bigskip
  \item[\texttt{revealRevocationKeyLender}(\(\mathsf{offer\_id},PRK_l\))]~\\[0.3em] 
  Purpose: This function enables the \textbf{lender} to disclose \(PRK_l\), which is the final step to formally complete the current installment and potentially the entire loan.

  \medskip
  Internal Steps:
  \begin{enumerate}[label=\arabic*., left=1em]
    \item \textbf{Parameter Validation:} The function first validates all provided input parameters. This includes:
    \begin{itemize}[nosep]
        \item ensuring that the loan offer identified by \texttt{positionId} currently holds an \textit{Opened} status.
        \item confirming that the \texttt{msg.sender} is indeed the \textbf{lender} associated with positionId.
        \item verifying that the status of the current installment is \textit{BorrowerRevocationKey}.
        \item checking that this function call occurs before the deadline of the current installment.
    \end{itemize}

    \item \textbf{Installment and Offer State Update:} The function updates the status of the current installment to \textit{LenderRevocationKey} and stores the provided \(PRK_l\) within the offer's state. If this particular installment is determined to be the final one in the loan sequence, the overall status of the loan offer is then updated to \textit{Successful}.
    \item \textbf{Fund Transfer:} The contract transfers the \texttt{installmentAmount} to the lender's address.
    \item \textbf{Event Emission:} A \texttt{LenderRevocationKeyRevealed}(offer\_id, installment\_id) event is emitted.
  \end{enumerate}

\end{description}

\chapter{Single-sided liquidity aggregation}
\textit{WIP}

\section{Staker Registration \& TSS Recomposition}
We can extend the basic loan‐channel construction by introducing a dynamic set of \emph{stakers} who jointly manage the \emph{liquidity pool} via a threshold signature. Let:
\begin{itemize}
  \item $\mathcal{A}$: active staker list, $|\mathcal{A}|\le N$.
  \item $stake_{\min}$: minimum stake amount for participation.
  \item $\Delta$: base increment for computing stake levels.
  \item $coef_k$: coefficient for the $k$‐th staker such that $stake_k = stake_{\min} + coef_k \cdot \Delta$.
  \item $T_0$: global timelock after which new stakers may no longer join.
  \item $t_{\mathit{epoch}}$: duration of one epoch for join/leave requests.
  \item $\Delta t$: duration of the TSS recomposition window at each epoch’s end.
\end{itemize}

During each epoch:
\begin{enumerate}[label=\roman*)]
  \item Collect join/leave requests; maintain a waiting list sorted descending by stake.
  \item While $|\mathcal{A}|<N$, move the highest waiting candidate into $\mathcal{A}$ (locking their stake).
  \item If $|\mathcal{A}|=N$ and new candidates exist, compare the smallest $stake_i$ in $\mathcal{A}$ with the highest in the waiting list; if the new stake is larger, evict the least staker and promote the candidate.
\end{enumerate}

At the end of each epoch, the contract initiates a TSS recomposition round over $\mathcal{A}$:
\begin{enumerate}[label=\roman*)]
  \item Sort $\mathcal{A}$ by decreasing stake.
  \item Request each staker to participate in the TSS DKG and signing process.
  \item If a staker fails to respond within $\Delta t$:
    \begin{itemize}
      \item Identify the non-responder and move them to the waiting list (unlocking their funds).
      \item Promote the top waiting candidate into $\mathcal{A}$ (locking their stake).
      \item Recompute all weights
      \[
        w_i = \frac{stake_i}{\sum_{j\in\mathcal{A}}stake_j},
        \quad i=1,\dots,|\mathcal{A}|,
      \]
      and rerun the DKG until a full set of $N$ participants completes.
    \end{itemize}
  \item When $N$ responsive stakers produce the TSS key, finalize the new aggregate key and record the weights.
\end{enumerate}

\section{Loan Acceptance \& Channel Creation}
When the borrower decides to take a loan (before $T_0$), they open a Lightning-style channel. The funding outputs to the $n=|\mathcal{A}|$ stakers are split according to their stakes:
\begin{align*}
  \mathsf{TX}_{{\rm comm}_0}^B &:~\{(\mathsf{TX_{fund}},0,(-,-));~(0\bitcoin,\,(addr_B\land LT_B)\lor(P_{\mathrm{agg}}\land P_B^{{\rm rev}_0})),\\
       &\quad (a_1\bitcoin,addr_{L_1}),\dots,(a_n\bitcoin,addr_{L_n})\},\\
  \mathsf{TX}_{{\rm comm}_0}^L &:~\{(\mathsf{TX_{fund}},0,(-,-));~(a_1\bitcoin,\,(addr_{L_1}\land LT_{L_1})\lor(P_B\land P_{L_1}^{{\rm rev}_0})),\\
       &\quad \dots,~(a_n\bitcoin,\,(addr_{L_n}\land LT_{L_n})\lor(P_B\land P_{L_n}^{{\rm rev}_0})),~(0\bitcoin,addr_B)\},
\end{align*}
where each
\[
  a_i = \frac{stake_i}{\sum_{j\in\mathcal{A}}stake_j}\cdot a,
  \quad i=1,\dots,n,
\]
and $a_i$ is decremented proportionally with each borrower repayment.

\section{Staker Participation Penalty Mechanism}
Each staker $i\in\mathcal{A}$ maintains a failure counter $f_i\in\mathbb{N}$, initially zero.  At every TSS recomposition or signing epoch:
\begin{enumerate}
  \item On timely valid response, set $f_i\leftarrow0$.
  \item Otherwise, $f_i\leftarrow f_i+1$:
    \begin{itemize}
      \item If $f_i=1$, fine 10\% of their stake: $stake_i\leftarrow0.9\,stake_i$.
      \item If $f_i\ge2$, remove $i$ from $\mathcal{A}$ and forfeit remaining $stake_i$.
    \end{itemize}
\end{enumerate}

\begin{algorithm}[H]
\caption{Enforcing Participation Penalties}
\begin{algorithmic}[1]
\Require Active set $\mathcal{A}$ with $\{stake_i,f_i\}$
\ForAll{$i\in\mathcal{A}$}
  \If{valid response from $i$}
    \State $f_i\leftarrow0$
  \Else
    \State $f_i\leftarrow f_i+1$
    \If{$f_i=1$}
      \State $stake_i\leftarrow0.9\times stake_i$  \Comment{10\% fine}
    \ElsIf{$f_i\ge2$}
      \State remove $i$; forfeit $stake_i$
    \EndIf
  \EndIf
\EndFor
\end{algorithmic}
\end{algorithm}

\section{Multi-Borrower Support}
To accommodate sequential borrowers $B_1, B_2,\dots$, the same \texttt{LoanContract} stores a mapping of channel states per borrower.  Upon borrower $B_j$ drawing amount $b_j$:
\[
  \text{for each }i\in\mathcal{A}:\quad
  stake_i\leftarrow stake_i - \frac{stake_i}{\sum_{k\in\mathcal{A}}stake_k}\cdot b_j,
\]
then:
\begin{enumerate}[label=\roman*)]
  \item Remove all $i$ with $stake_i<stake_{\min}$ (unlock their funds).
  \item At epoch end, accept new stakers as usual using the updated $stake_i$ values.
  \item Recompose the TSS key using the epoch procedure over the updated $\mathcal{A}$.
\end{enumerate}

In the base protocol, the loan contract provides a single \texttt{registerChannel} entry point:
\[
  \texttt{registerChannel}(TX_{\tt fund},\,\pi_{\mathsf{SPV}},\,LT_B,\,LT_L)
\]
which locks \((b + c)\) USD and records one borrower’s channel. To allow multiple borrowers \(B_1, B_2,\dots\) to draw sequential loans from the same contract, we simply generalize:

\begin{itemize}
  \item Internally maintain a list (or mapping) of \(\mathit{ChannelState}_j\) for each borrower \(B_j\).
  \item The same \texttt{registerChannel} function appends a new \(\mathit{ChannelState}_j\) rather than blocking further calls.
  \item All funding TX, SPV proof, locktimes, principal \(b_j\), deposit \(c_j\) are stored under \(j\).
\end{itemize}
No separate contract is required: one \texttt{LoanContract} can safely host all sequential borrowers, provided it enforces that each \(\sum_j b_j\le\Sigma\) (the total available USD) and manages each channel’s lifecycle independently.

\subsection{On-chain Call Flow for Borrower \(B_j\)}  
\label{sec:multi-borrower-flow}

\begin{enumerate}
  \item \texttt{requestLoan}(\(B_j,a_j,T0_j,TX_{\tt fund}^j,TX_{\tt comm0}^j\)):  
    borrower \(B_j\) pays no on-chain value; they register their pre-agreed funding and initial commitment TXs.
  \item \texttt{acceptLoan}(\(\,\sigma_L^j, c_j\)):  
    lender confirms by providing the signed channel commitment and locking \((b_j + c_j)\) USD.
  \item \texttt{finalizeLoan}(\(\sigma_B^j,\pi_{\mathsf{SPV}}(TX_{\tt fund}^j)\)):  
    borrower proves funding on Bitcoin and receives \(b_j\) USD.
\end{enumerate}

Each of these calls reads/writes only that borrower’s \(\mathit{ChannelState}_j\), so multiple borrowers can coexist.

\section{Signer Identification during TSS Key Generation}

To precisely identify non-responders (malicious or offline) during the TSS DKG for public key creation, we record each phase’s contributions.  Let \(\mathcal{M}\) be the set of identified non-responders.

\begin{algorithm}[H]
\caption{Identifying Malicious Stakers in TSS DKG}
\begin{algorithmic}[1]
\Require Active set \(\mathcal{A}\)
\State \(\mathcal{M}\gets\emptyset\)  \Comment{Initialize malicious set}
% Phase 1: Commitments
\ForAll{\(i\in\mathcal{A}\)}
  \If{commitment\(_i\) not received or invalid}
    \State \(\mathcal{M}\gets\mathcal{M}\cup\{i\}\); remove \(i\) from \(\mathcal{A}\)
  \Else
    \State record valid commitment\(_i\)
  \EndIf
\EndFor
% Phase 2: Share distribution
\ForAll{\(i\in\mathcal{A}\)}
  \If{share\(_i\) not received or fails verification}
    \State \(\mathcal{M}\gets\mathcal{M}\cup\{i\}\); remove \(i\) from \(\mathcal{A}\)
  \Else
    \State record valid share\(_i\)
  \EndIf
\EndFor
% Demotion and refill
\ForAll{\(i\in\mathcal{M}\)}
  \State move \(i\) to waiting list, take fine 
  \State promote top candidate into \(\mathcal{A}\); lock their stake
  \State recompute weights and rerun DKG until \(|\mathcal{A}|=N\)
\EndFor
\State Finalize TSS public key over remaining \(\mathcal{A}\)
\end{algorithmic}
\end{algorithm}

\chapter{Authors and acknowledgments}
This write-up was prepared by Oleksandr Kurbatov, Kyrylo Baibula, Yaroslava Chopa, Sergey Kozlov, and Oleh Komendant.

For the first implementation of the SPV contract and prover, as well as a toy implementation of the Wrapless protocol, we would also like to acknowledge Illia Dovgopoly, Dmitrii Kurbatov, Zakhar Naumets, and Yulia Artikulova. 

And of course, it is difficult to overrate the valuable feedback and proposals from Pavel Kravchenko, Vlad Dubinin, Lasha Antadze, Yaroslav Panasenko, and Mykhailo Velykodnyi. 
\printbibliography

\end{document}